%% file: MU_Codes_Journal_DC_FINAL.tex
\documentclass[twocolumn,twoside]{IEEEtran}
\usepackage[utf8]{inputenc}
\usepackage{graphicx}
\usepackage{amssymb}
\usepackage[cmex10]{amsmath}
\usepackage{color}
\usepackage{theorem}
\usepackage{hyperref}
\usepackage{url}
\usepackage{breakurl}
\usepackage{booktabs}
\usepackage{bbm}
\usepackage{bbold}
\usepackage{comment}

\usepackage{multirow}
\usepackage{booktabs}
\usepackage{float}

\usepackage[english]{babel}
\usepackage[english = american]{csquotes}
\usepackage{mathtools}
\usepackage{mathdots}
\MakeOuterQuote{"} 

\definecolor{light_gray}{rgb}{0.6,0.6,0.6}
\definecolor{awgray}{rgb}{0.7,0.7,0.7}
\definecolor{awgray_dark}{rgb} {0.4,0.4,0.4}

\usepackage{pgf}
\usepackage{tikz}
\usetikzlibrary{matrix,chains,positioning,arrows,calc,decorations,plotmarks,patterns, fit,backgrounds}
\usepackage{pgfplots}
\usepgfplotslibrary{external}
\usetikzlibrary{external}                       
\tikzsetexternalprefix{tikzpic/}                   

\pgfplotscreateplotcyclelist{mylist}{red,blue,black,yellow,brown}

\tikzset{
    >=stealth',
    mycircle/.style={circle, draw=black, thick, text width=.1em, minimum height=.8em, text centered, font=\scriptsize}, 
    mycircle_gray/.style={circle, draw=gray, thick, text width=.1em, minimum height=.8em, text centered, font=\tiny},    
    mycircle_small/.style={circle,draw=awgray_dark,fill = awgray_dark, inner sep=0,minimum size=.6em},       
    mycircle_small_black/.style={circle,draw=black,fill = black, inner sep=0,minimum size=.6em},   
    mybox/.style={rectangle,rounded corners,draw=black, thick,text width=1em,minimum height=4em,minimum width=4em,text centered},     
    mybox_small/.style={rectangle,rounded corners,draw=black, thick,text width=1em,minimum height=2em,minimum width=2em,text centered},               
    mybox_vec/.style={rectangle,rounded corners,draw=black, thick,text width=1em,minimum height=0.7em, minimum width=4em,text centered},  
    mybox_vec_short/.style={rectangle,rounded corners,draw=black, thick,text width=1em,minimum height=0.7em, minimum width=2em,text centered},                  
    pil/.style={->, thick, shorten <=2pt, shorten >=2pt,},
}

\usepackage{algorithm,algpseudocode}

\makeatletter
\newcommand{\removelatexerror}{\let\@latex@error\@gobble}
\makeatletter
\newcommand{\proofpart}[2]{%
	\par
	\addvspace{\medskipamount}%
	\noindent\emph{Part #1: #2}\par\nobreak
	\addvspace{\smallskipamount}%
	\@afterheading
}
\makeatother

\newtheorem{innercustomgeneric}{\customgenericname}
\providecommand{\customgenericname}{}
\newcommand{\newcustomtheorem}[2]{%
	\newenvironment{#1}[1]
	{%
		\renewcommand\customgenericname{#2}%
		\renewcommand\theinnercustomgeneric{##1}%
		\innercustomgeneric
	}
	{\endinnercustomgeneric}
}
\newcustomtheorem{customproposition}{Proposition}
\newcustomtheorem{customtheorem}{Theorem}
\newcustomtheorem{customlemma}{Lemma}
\newcustomtheorem{customclaim}{Claim}
\IEEEoverridecommandlockouts

\input{mydefs.tex}

\begin{document}
\title{Mutually Uncorrelated Codes for DNA Storage}

\author{\large Maya~Levy and Eitan~Yaakobi,~\IEEEmembership{Senior Member,~IEEE} 
\thanks{M. Levy and E. Yaakobi are with the Department of Computer Science, Technion --- Israel Institute of Technology, Haifa 32000, Israel (e-mail: \texttt{\{mayalevy,yaakobi\}@cs.technion.ac.il}).
}
\thanks{This work was supported in part by the Israel Science Foundation under Grant 1624/14. Part of the results in the paper were submitted to the IEEE International Symposium on Information Theory, Aachen, Germany, June 25 -- 30, 2017 (reference~\cite{LY17}).}}

\maketitle

\begin{abstract}
	\emph{Mutually Uncorrelated} (\emph{MU}) codes are a class of codes in which no proper prefix of one codeword is a suffix of another codeword. These codes were originally studied for synchronization purposes and recently, Yazdi et al. showed their applicability to enable random access in DNA storage. In this work we follow the research of Yazdi et al. and study MU codes along with their extensions to correct errors and balanced codes. 
	We first review a well known construction of MU codes and study the asymptotic behavior of its cardinality. This task is accomplished by studying a special class of run-length limited codes that impose the longest run of zeros to be at most some function of the codewords length. We also present an efficient algorithm for this class of constrained codes and show how to use this analysis for MU codes. 
	Next, we extend the results on the run-length limited codes in order to study $(d_h,d_m)$-MU codes that impose a minimum Hamming distance of $d_h$ between different codewords and $d_m$ between prefixes and suffixes. In particular, we show an efficient construction of these codes with nearly optimal redundancy. We also provide similar results for the edit distance and balanced MU codes. Lastly, we draw connections to the problems of comma-free and prefix synchronized codes. \end{abstract}
\begin{IEEEkeywords}
	DNA storage, mutually uncorrelated codes, constrained codes, non-overlapping codes, cross--bifix-free codes, comma-free codes.
\end{IEEEkeywords}

\section{Introduction}\label{sec:introduction}
\renewcommand{\baselinestretch}{0.94}\normalsize
\emph{Mutually Uncorrelated} (\emph{MU}) codes satisfy the constraint in which the prefixes set and suffixes set of all codewords are disjoint. This class of codes was first studied by Levenshtein~\cite{L64} for the purpose of synchronization, and has received attention recently due to its relevance and applicability for DNA storage~\cite{Y15}. Namely, these codes offer random access of DNA blocks in synthetic DNA storage.

The potential of DNA molecules as a volume for storing data was recognized due to its unique qualities of density and durability.
The first large scale DNA storage system was designed by Church et al.~\cite{C12} in 2012. Since then a few similar systems were implemented for archival applications as they did not support random access to the memory.
Recently, in~\cite{B16, Y15} two random access DNA storage systems were proposed. To enable random access, the authors suggested to equip the DNA information blocks with unique addresses, also known as \emph{primers}. Under this setup, the reading process starts with a phase aimed to identify the requested DNA block. During the identification process the complementary sequence of the unique primer is sent to the DNA pool and stitches to the primer part of the requested DNA block. By that, the selection of the DNA block is done and only the selected block is read.

Obtaining a \emph{good} set of primers is therefore a key property in achieving the desired random access feature as it guarantees the success of the chemical processes involved in the identification phase. 
The constraints that a good primers set should satisfy are listed by Yazdi et al. in~\cite{Y15}. In this paper we focus on three of the four listed constraints, described as follows: 1) the MU constraint is imposed as it avoids overlaps in two primers, which are likely to cause erroneous identification of DNA blocks, 2) both the writing and reading channels of DNA introduce substitution errors, therefore we are interested in large mutual Hamming distance, and 3)  we require the primers to be balanced since balanced DNA sequences increase the chances of successful reads.
In addition, the authors of~\cite{YK15} mentioned that deletion errors are introduced during synthesis. We therefore also extend our study to MU codes with edit distance. 

MU codes were rigorously studied in the literature under different names such as \emph{codes without overlaps}~\cite{L70,L04}, \emph{non-overlapping codes}~\cite{ B15} and \emph{cross-bifix-free codes}~\cite{BA16, B12, C13}. However, the basic problem of finding the largest MU code is still not fully solved. Let us define by $A_{MU}(n,q)$ the size of the largest MU code over a field of size $q$. The best upper bound, found by Levenshtein~\cite{L70}, states that $A_{MU}(n,q)< q^n/(e(n-1))$, while the best known constructive lower bound, given independently in~\cite{C13,G60,L64}, states that $A_{MU}(n,q)\gtrsim  \frac{q-1}{qe}\cdot \frac{q^n}{n}$. Hence, for the binary case there is still a gap of factor 2 between the lower and upper bounds. The construction of MU codes is explicit. Given some $k<n$, one fixes the first $k$ bits to be zero, followed by a single one, and the last bit is one as well. The sequence of the remaining $n-k-2$ symbols needs to satisfy the constraint that it does not have a zeros run of length $k$. Previous results claimed that $\frac{q-1}{qe}\cdot \frac{q^n}{n}$ is a lower bound on the construction's code size, when $n=(q^i-1)/(q-1)$, however it was not known whether it is possible to achieve codes with larger cardinality. We give an explicit expression of the asymptotic cardinality of these codes for any value of $n$ and show that the lower bound $\frac{q-1}{qe}\cdot \frac{q^n}{n}$ is indeed tight. 
This result is accomplished by studying a special family of run-length limited constrained codes, called \emph{$k$-run length limited} (\emph{RLL}) \emph{codes}, which impose the longest run of zeros to be of size at most $k-1$, where $k$ is a function of the word's length. We also present an efficient encoding and decoding algorithm for $k$-RLL codes with linear complexity which results in efficient binary MU codes with $\lceil \log(n)\rceil +4$ redundancy bits.

Next, we extend the study of $k$-RLL codes to the \emph{window weight limited constraint} that imposes the Hamming weight of every length-$k$ subsequence to be at least some prescribed number $d$. Accordingly, we study MU codes with error-correction capabilities. A \emph{$(d_h,d_m)$-MU code} is an MU code with minimum Hamming distance $d_h$, with the additional property that every prefix of length $i$ differs by at least $\min\{i,d_m\}$ symbols from all proper length suffixes. We show an upper bound on the size of $(d_h,d_m)$-MU codes and give a construction of such codes with encoder and decoder of linear complexity. The redundancy of the construction is $\left\lfloor\frac{d_h+1}{2}\right\rfloor\log(n) + (d_m-1) \log\log n + \mathcal{O} (d_m \log d_m)$, which is nearly optimal with respect to our upper bound on the code size. A similar constraint is imposed when studying MU codes with edit distance. We  give a general result of such codes and study MU codes that can correct a single deletion or insertion. For the latter case we use a systematic encoder for the Varshamov Tenengolts codes~\cite{VT65}. Lastly, we study balanced MU codes, that is, codes which are both MU and balanced. We show that the achievable minimum redundancy of these codes is approximately $1.5\log(n)$ and for an efficient construction we use Knuth's algorithm while the redundancy is roughly $2\log(n)$ bits.

The rest of this paper is organized as follows. In Section~\ref{sec:defs}, we formally define the codes studied in the paper and review related work. In Section~\ref{sec:RLL}, we analyze the redundancy of $k$-RLL codes, when $k$ is a function of the word's length, and propose efficient encoding and decoding algorithms for these codes. We use this analysis in Section~\ref{sec:MUCodes} in order to study the asymptotic cardinality of the MU codes. We extend the results on $k$-RLL codes in Section~\ref{sec:WWL} to study the window weight limited constraint and accordingly in Section~\ref{sec:ECC_MUCodes}, we extend the class of MU codes to $(d_h,d_m)$-MU codes. 
We continue in Section~\ref{MUEdit} to study MU codes that can correct deletions and insertions, and in Section~\ref{BalancedMUCodes} we study balanced MU codes. Furthermore, in Section~\ref{sec:other}  we draw connections to the problems of comma-free and prefix synchronized codes. Lastly, Section~\ref{sec:conc} concludes and summarizes the results in the paper. 

\section{Definitions, Preliminaries, and Related Work}\label{sec:defs}

For every two integers $i\le k$ we denote by $[i,k]$ the set of integers $\{j \ | \ i\le j \le k\}$ and use $[k]$ as a shortening to $[1,k]$. We use the notation of $\Sigma= \{0,1\}$ as the binary alphabet and $\Sigma_q = \{0,1,\ldots,q-1\}$ is the notation for larger alphabets. For two integers $n,k,$ the notation $\langle n \rangle _k $ stands for $n$ modulo $k$. For a vector $\vec{a}=(a_1,\ldots,a_n)$ and $i,j \in [n], i\le j$, we denote by $\vec{a}_i^j$ the subvector $(a_i,\ldots,a_j)$ of $\vec{a}$. For $j<i$, $\vec{a}_i^j$ is the empty word. 
The Hamming weight of a vector $\vec{a}$ is denoted by $w_H(\vec{a})$ and $d_H(\vec{a},\vec{b})$ is the Hamming distance between $\vec{a}$ and $\vec{b}$. A \emph{zeros run} of length $r$ of a vector $\vec{a}$ is a subsequence $\vec{a}_i^{i+r-1}$, $i \in [n-r+1]$ such that $a_i = \cdots = a_{i+r-1} = 0$. The notation $\vec{a}^i$ denotes the concatenation of the vector $\vec{a}$ $i$ times and $\vec{a}\vec{b}$ is the concatenation of the two vectors $\vec{a}$ and $\vec{b}$. Let $A,B$ be two sets of vectors over $\Sigma_q$. We denote the set $AB = \{ab|a\in A, b\in B\}$ and the set $A^i=AA\cdots A$, to be $i$ concatenations of the set $A$. For two functions $f(n), g(n)$ we say that $f(n)\lesssim g(n)$ if $\lim_{n\rightarrow \infty} {\frac{f(n)}{g(n)}} \le 1$, and $f(n)\approx g(n)$ if $\lim_{n\rightarrow \infty} {\frac{f(n)}{g(n)}} =1 $.  
The redundancy of a set $A\subseteq \Sigma_q^n $ is defined as $\textmd{red}(A)=n - \log_q |A|$. If the base of the logarithm is omitted then it is assumed to be 2.

\begin{definition}
Two not necessarily distinct words $\vec{a},\vec{b}\in \Sigma_q^n$ are \textbf{mutually uncorrelated} if any non-trivial prefix of $\vec{a}$ does not match any non-trivial suffix of $\vec{b}$. A code $\cC \subseteq \Sigma_q^n$ is a \textbf{mutually uncorrelated} (\textbf{MU}) \textbf{code} if any two not necessarily distinct codewords of $\cC$ are mutually uncorrelated. 
\end{definition}

Let us denote by $A_{MU}(n,q)$ the largest cardinality of an MU code of length $n$ over $\Sigma_q$. Levenshtein showed in~\cite{L70} that for all $n$ and $q$ 
\begin{equation}\label{eq:ub}
A_{MU}(n,q) \le \left( \frac{n-1}{n} \right)^{n-1}\frac{q^n}{n} < \frac{q^n}{e(n-1)},
\end{equation}
or in other words, the redundancy is lower bounded by $\log e + \log (n-1)$.  

We now recall a well studied family of MU codes. It was suggested independently first by Gilbert in~\cite{G60}, later by Levenshtein in~\cite{L64} and recently by Chee et al. in~\cite{C13}.
\begin{construction}\label{const:mu}
	Let $n,k$ be two integers such that $1\le k < n$ and let $\cC_1(n,q,k)\subseteq \Sigma_q^n$ be the following code:
	\begin{align*}
	\cC_1(n,q,k)=\{\vec{a}\in \Sigma_q^n \ | & \ \forall i\in [k], a_i = 0, a_{k+1}\neq0, a_n \ne 0, \\
	&	 \vec{a}_{k+2}^{n-1} \ \textmd{has no zeros run of length} \ k \}.
	\end{align*}
\end{construction}
We denote $$\cC_1(n,q) = \max_{1\le k < n} {\{|\cC_1(n,q,k)|\}}.$$
It was proved in~\cite{C13,G60,L64} that there exists an appropriate choice of $k$ for which the following lower bounds hold 
\begin{equation} \label{eq:mulb1}
\cC_1(n,q) \gtrsim q^{-\frac{q}{q-1}}\ln q \frac{q^n}{n},
\end{equation}
as $n\rightarrow \infty$, and more specifically
\begin{equation} \label{eq:mulb2}
\cC_1(n,q) \gtrsim \frac{q-1}{qe}\cdot \frac{q^n}{n},
\end{equation}
where $n\rightarrow \infty$ over the subsequence $\frac{q^i-1}{q-1}, i\in \mathbb{N}$. However, it was not established in these works what the asymptotic cardinality of $\cC_1(n,q)$ is and for which $k$ it is achieved. We answer these questions in Section~\ref{sec:MUCodes} and show that the asymptotic inequalities in~(\ref{eq:mulb1}), ~(\ref{eq:mulb2}) are asymptotically tight.  

Additional interesting approach for constructing binary MU codes was proposed in~\cite{B12} and was later generalized in~\cite{BA16} to alphabets of size $q>2$. In this case, it was shown that for some small values of $n$ the approach in~\cite{BA16} achieves larger cardinality than Construction \ref{const:mu}. Both works~\cite{BA16,B12} do not offer asymptotic analysis of the constructions' redundancy, however, for the binary case it is commonly believed that Construction \ref{const:mu} provides the best known asymptotic code size. Furthermore, the author of~\cite{B15} extended Construction~\ref{const:mu} for $q>2$ and showed that if $q$ is a multiple of $n$ then this extension results with strictly optimal codes. Lastly, we note that in~\cite{B14,B17} Gray codes were presented for listing the vectors of the code $\cC_1(n,q,k)$. 

Due to the structure of Construction~\ref{const:mu}, the tools that we use for analyzing its cardinality are taken from studies in the field of constraint coding. 
We recall the well known Run Length Limited (RLL) constraint in which the lengths of the zero runs are limited to a fixed range of values. 

\begin{definition}
	Let $k$ and $n$ be two integers. We say that a vector $\vec{a}\in \Sigma_q^n$ satisfies the $k$\textbf{-run length limited (RLL)} constraint, and is called a $k$\textbf{-RLL vector}, if $n<k$ or $\forall i\in [n-k+1]: w_H(\vec{a}_i^{i+k-1})\ge 1$.
\end{definition}
In other words, a vector is called a $k$-RLL vector if it does not contain a zeros run of length $k$.
We denote by $A_q(n,k)$ the set of all {$k$-RLL} length-$n$ vectors over the alphabet $\Sigma_q$, and $a_q(n,k) = |A_q(n,k)|$.

The definition of the $k$-RLL constraint is similar to the $(d,k)$-RLL constraint from~\cite{M01} that states that any zeros run is of length at least $d$ and at most $k$. In other words, the $k$-RLL constraint is equivalent to the well studied $(0,k-1)$-RLL constraint. We chose the notation of $k$-RLL for simplicity of the following analysis to come. 

The capacity of the $k$-RLL constraint, is defined for any fixed values of $k$ and $q$ as
$$E_{k,q}=\lim_{n\rightarrow\infty}\frac{\log(a_q(n,k))}{n}.$$ 
Kato and Zeger~\cite{KZ05} provided the following result for the binary case and later Jain et. al generalized it for $q\ge 2$ in \cite{J17} 
\begin{equation*} \label{Jain}
E_{k,q}=\log q - \frac{(q-1) \log e}{ q^{k+2}}(1 + o(1)).
\end{equation*}

\section{Run Length Limited Constraint}\label{sec:RLL}
The $k$-RLL constraint was studied extensively in the literature previously by many works; see~\cite{M01} and references therein. However, this study was focused solely for the case in which $k$ is fixed, meaning $k$ is independent of $n$. As it will be explained later, the MU codes problem requires analysis of values of $k$ which are dependent on $n$. In Subsection~\ref{subsec:cardinality}, we resort to previous results on the $k$-RLL constraint, when $k$ is fixed to provide a better understanding on the asymptotic behavior of $a_q(n,k)$ when $k$ depends on $n$. Later in Subsection~\ref{subsec:encoding} we present efficient encoding and decoding algorithms to avoid zeros runs of length $\lceil \log n \rceil +1$, with linear time and space complexity, and with only a single bit of redundancy.
\subsection{Cardinality Analysis} \label{subsec:cardinality}
 We start by giving general bounds on $a_q(n,k)$ for most values of $n,k$ in Lemma~\ref{lem:bounds} and Lemma~\ref{lem:cardinality}. The intuition of the proofs is as follows. In Lemma~\ref{lem:bounds}, we consider the set of all length-$n$ vectors which are a concatenation of multiple length-$2k$ $k$-RLL vectors, i.e. each length-$2k$ vector satisfies the $k$-RLL constraint. The set $A_q(n,k)$ is a subset of this set and hence $a_q(n,k)$ is upper bounded by its size. For the lower bound we consider the set of all vectors of length $n$ which are again, a concatenation of multiple shorter $k$-RLL vectors, however, now we extract vectors that start or end with $\lceil k/2 \rceil$ zeros. The new set is a subset of $A_q(n,k)$ and a lower bound is derived accordingly.
 
 In Lemma~\ref{lem:cardinality} we study the value $a_q(mn,k)$ and bound it similarly to the bounds in Lemma~\ref{lem:bounds} by considering a concatenation of $m$ vectors of length $n$. Both the upper and lower bounds on $a_q(mn,k)$ will include the size $a_q(n,k)$ and thus equivalent upper and lower bounds on $a_q(n,k)$ will be established. We then derive the final bounds using the expression $E_{k,q}= \lim_{m\rightarrow\infty}\frac{\log(a_q(mn,k))}{mn}$ for the capacity. 

\begin{lemma} \label{lem:bounds} 
	Let $n,k$ be  positive integers such that $5\le k\le n,$ then
	$$q^n(\frac{q-1}{q})^{c_2\frac{n}{q^{k-1}}}e^{-\frac{c_2}{c_1}\frac{n}{q^{1.5k-1}}}\le a_q(n,k) \le q^{n- c_3\frac{n-2k}{q^k}} , $$ 
	where $c_1=\frac{(q-1)q^{\lceil{k/2}\rceil-k/2}}{2q},c_2=\frac{\lfloor\frac{n}{q^{k-1}}\rfloor+1}{\frac{n}{q^{k-1}}}, c_3=\frac{\log_q e (q-1)^2}{2q^{2}}$.
	
\end{lemma}
\begin{IEEEproof}
	\proofpart{1}{Upper Bound}
	We consider the set $A_q(2k,k)^{\lfloor\frac{n}{2k}\rfloor}$, that is, the set of vectors which are a concatenation of $\lfloor\frac{n}{2k}\rfloor$ vectors from $A_q(2k,k)$. We then append it with the set of all length-$\langle n\rangle_{2k}$ $q$-ary vectors. The resulting set of length-$n$ vectors is denoted by $$B_q(n,k) =A_q(2k,k)^{\lfloor\frac{n}{2k}\rfloor}\Sigma_q^{\langle n\rangle_{2k}}.$$
	Note that $A_q(n,k) \subseteq B_q(n,k)$ and $|B_q(n,k)|=a_q(2k,k)^{\lfloor\frac{n}{2k}\rfloor}q^{\langle n\rangle_{2k}}$ , hence
	\begin{equation} \label{eq:ank}
	a_q(n,k)\le a_q(2k,k)^{\lfloor\frac{n}{2k}\rfloor}q^{\langle n\rangle_{2k}}.
	\end{equation} 
	
	Let $b(k)$ be the number of vectors of length $2k$ with a zeros run of length  exactly $k$ and with no zeros run of length greater than $k$. There are $q^{2k-(k+1)}(q-1)$ vectors of length $2k$ that start with a zeros run of length exactly $k$ and $q^{2k-(k+1)}(q-1)$ different vectors that end with such a run. There are $2k-(k-1)-2=k-1$ other positions in which a zeros run of length $k$ can start within the $2k$ vector. For each such a position there are $q^{2k-(k+2)}(q-1)^2$ different vectors with a zeros run of length exactly $k$. In total we have 
	\begin{align*}
	b(k) &= 2 q^{2k-(k+1)}(q-1)+(k-1)q^{2k-(k+2)}(q-1)^2\\
	&= 2 q^{k-1}(q-1)+(k-1)q^{k-2}(q-1)^2 \\
	&\ge (k+1)q^{k-2}(q-1)^2.
	\end{align*}
	All length-$2k$ vectors with a zeros run of length exactly $k$ are not included in $A_q(2k,k)$. Therefore,
	\begin{equation} \label{eq:a2k}
	a_q(2k,k) \le q^{2k} - b(k) \le q^{2k} - (k+1)q^{k-2}(q-1)^2.
	\end{equation} 
	
	By combining inequalities (\ref{eq:ank}) and (\ref{eq:a2k}), we get 
	\begin{align*}
	a_q(n,k) & \le (q^{2k} - (k+1)q^{k-2}(q-1)^2)^{\lfloor\frac{n}{2k}\rfloor}q^{\langle n\rangle_{2k}}   \\ 	
	& = (q^{2k}(1-\frac{(k+1)(q-1)^2}{q^{k+2}}))^{\lfloor\frac{n}{2k}\rfloor}q^{\langle n\rangle_{2k}}   \\ 	
	& = q^n(1-\frac{(k+1)(q-1)^2}{q^{k+2}})^{\lfloor\frac{n}{2k}\rfloor}   \\ 	
	& \stackrel{(a)}{\le}q^n(e^{-\frac{(k+1)(q-1)^2}{q^{k+2}}})^{\lfloor\frac{n}{2k}\rfloor}   \\ 
	& \le q^{n-\log_q e \frac{(k+1)(q-1)^2}{q^{k+2}}(\frac{n}{2k}-1)}   \\ 	
	& \le q^{n-\log_q e \frac{(q-1)^2(n-2k)}{2q^{k+2}}},
	\end{align*}
	where (a) results from the inequality $1-x\le e^{-x} $ for all $x$. 

	\proofpart{2}{Lower Bound}
	We start by giving an upper bound on the number of length-$n$ vectors with a zeros run of length at least $k$. There are $n-k+1$ positions in which a zeros run can start and for each position we have at most $q^{n-k}$ different vectors. From the union bound we have that the number of length-$n$ vectors with a zeros run of length at least $k$ is upper bounded by $q^{n-k}(n-k+1)$, and by excluding those vectors from the set of all length-$n$ vectors we get 
	$$a_q(n,k)\ge q^n-q^{n-k}(n-k+1).$$
	This bound is irrelevant for values of $k$ smaller than $\log_q n$ as it is less than zero. However, if we choose $k\ge \log_q n+1$ it becomes useful and we get
	\begin{align*}
	a_q(n,k)&\ge q^n-q^{n-\log_q n-1}(n-\log_q n )\\
	&= q^n(1-\frac{n-\log_q n }{qn}) \\ 
	&\ge q^n(1-\frac{n }{qn}) \\ 
	&= q^{n-1}(q-1). 
	\end{align*}
	Since we are also interested in the case of $k<\log_q n+1$, or equivalently $n> q^{k-1}$, we continue our analysis by breaking down the length-$n$ vector to blocks of length $q^{k-1}$ and applying the bound on them in the following manner.
	    
	For $\ell \ge k$ we denote the set 
	\begin{align*}
	C_q(\ell,k) = \{ \vec{a} \mid \  &\vec{a}\in A_q(\ell,k), \\
	&\textmd{$\vec{a}$ starts or ends with $\lceil{k/2}\rceil$ zeros}    \}
	\end{align*}

	and $c_q(\ell,k) = |C_q(\ell,k)|$.
	Note that $$c_q(\ell,k)\leq 2\cdot q^{\ell-\lceil{k/2}\rceil}.$$ 
	We also denote $D_q(\ell,k) = A_q(\ell,k)\setminus C_q(\ell,k)$ and 
	\begin{align*}
	d_q(\ell,k) & = |D_q(\ell,k)| = a_q(\ell,k)- c_q(\ell,k) \\
	&\geq a_q(\ell,k) - 2q^{\ell-\lceil{k/2}\rceil}.
	\end{align*} 
	For $k\ge \log_q \ell+1$, or equivalently $\ell\le q^{k-1}$ we have that 
	\begin{align}
	d_q(\ell,k)&\ge q^{\ell-1}(q-1)-2q^{\ell-\lceil{k/2}\rceil} \nonumber\\ 
	&= q^{\ell-1}(q-1)(1-\frac{2q}{(q-1)q^{\lceil{k/2}\rceil}}). \label{eq:d_nk}
	\end{align} 
	
	Consider the set of vectors which are a concatenation of $\lfloor\frac{n}{q^{k-1}}\rfloor$ vectors from $D_q(q^{k-1},k)$ appended by a vector from $D_q(\langle n\rangle_{q^{k-1}},k)$. We denote this set by $$E_q(n,k)=D_q(q^{k-1},k)^{\lfloor\frac{n}{q^{k-1}}\rfloor}D_q(\langle n\rangle_{q^{k-1}},k).$$ Note that $E_q(n,k) \subseteq A_q(n,k)$ and $$|E_q(n,k)|=d_q(q^{k-1},k)^{\lfloor\frac{n}{q^{k-1}}\rfloor}d_q(\langle n\rangle_{2^{k-1}},k).$$ 
	Hence,
	\begin{align*}
	a_q(n,k)\ge & d_q(q^{k-1},k)^{\lfloor\frac{n}{q^{k-1}}\rfloor}d_q(\langle n\rangle_{q^{k-1}},k)\\
	\stackrel{Eq.(\ref{eq:d_nk})}{\ge}  &(q^{q^{k-1}-1}(q-1)(1-\frac{2q}{(q-1)q^{\lceil{k/2}\rceil}}))^{\lfloor\frac{n}{q^{k-1}}\rfloor}\\
	\cdot  &q^{\langle n\rangle_{q^{k-1}}-1}(q-1)(1-\frac{2q}{(q-1)q^{\lceil{k/2}\rceil}})\\
	= & q^n(\frac{q-1}{q})^{\lfloor\frac{n}{q^{k-1}}\rfloor+1}(1-\frac{2q}{(q-1)q^{\lceil{k/2}\rceil}})^{\lfloor\frac{n}{q^{k-1}}\rfloor+1}.
	\end{align*}
	For all $x<-1$, it is known that $(1+\frac{1}{x})^{x+1}<e$. We denote $x = -\frac{(q-1)q^{\lceil{k/2}\rceil}}{2q}=-c_1q^{\frac{k}{2}},$ for some constant $c_1>0$. For $k\ge 5, q\ge 2$ we have that $x<-1$ and we deduce that
	\begin{align*}
	(1-\frac{2q}{(q-1)q^{\lceil{k/2}\rceil}})^{\lfloor\frac{n}{q^{k-1}}\rfloor+1} =& (1+\frac{1}{x})^{(x+1)(\lfloor\frac{n}{q^{k-1}}\rfloor+1)/(x+1)}\\
	> & e^{(\lfloor\frac{n}{q^{k-1}}\rfloor+1)/(x+1)}\\
	\stackrel{(a)}{=} & e^{(c_2\frac{n}{q^{k-1}})/(-c_1q^{\frac{k}{2}})}\\
	= & e^{-\frac{c_2}{c_1}\frac{n}{q^{1.5k-1}}},
	\end{align*}
	where (a) follows from a choice of $c_2>0$ as the constant which satisfies $c_2\frac{n}{q^{k-1}} = \lfloor\frac{n}{q^{k-1}}\rfloor+1$. Finally, we conclude that
	\begin{align*}	
	a_q(n,k)\ge& q^n(\frac{q-1}{q})^{c_2\frac{n}{q^{k-1}}}e^{-\frac{c_2}{c_1}\frac{n}{q^{1.5k-1}}}.
	\end{align*}
\end{IEEEproof}	
From Lemma~\ref{lem:bounds} we have that 
\begin{align*}
c_3\frac{n-2k}{q^k} \le & \textmd{red}(A_q(n,k))\\
					\le &\log (\frac{q}{q-1})c_2\frac{n}{q^{k-1}}+ \frac{\log (e) c_2}{c_1}\frac{n}{q^{1.5k-1}}.
\end{align*}

If $n-2k=\Theta(n)$ the lower and upper bounds are of the same order, meaning there exist constants $C_1,C_2>0$ such that for large enough $n$
$$C_1\frac{n}{q^k} \le \textmd{red}(A_q(n,k))\le C_2\frac{n}{q^k} , $$ 
and the next corollary follows.
\begin{corollary} \label{cor:theta}
	Let $f(n)$ be a function such that $n-2(\log_q n -f(n))=\Theta(n)$ and $\log_q n -f(n)$ is a positive integer. Then, the redundancy of $A_q(n,\log_q n -f(n))$ is $\Theta(q^{f(n)}).$ 
\end{corollary}

Note that $f(n)$ can be negative. The result from Corollary~\ref{cor:theta} provides us with a general understanding on how the redundancy of the set $A_q(n,k)$ behaves for different values of $k$. We can conclude for example that for $k=0.5\log_q n$ the redundancy is $\Theta(\sqrt{n})$. Motivated by the MU problem, we are interested in further exploring the case in which the redundancy is constant. According to the result from Lemma~\ref{lem:bounds} and Corollary~\ref{cor:theta}, this holds when $f(n)$ is a constant function. That is, we are interested in the asymptotic behavior of $a_q(n,\lceil \log_q n\rceil + z), z\in \mathbb{Z}$ up to a better precision than the one suggested in Lemma~\ref{lem:bounds}. For this purpose, we provide in the next Lemma additional results on the asymptotic behavior of $a_q(n,k)$.
\begin{lemma}\label{lem:cardinality} Let $n,k$ be integers such that $0<k\le n$. Then,
	$$2^{nE_{k,q}}\leq {a_q(n,k)} \leq 2^{nE_{k,q}} + 2q^{n-\lceil{k/2}\rceil}.$$
\end{lemma}

\begin{IEEEproof}
	\proofpart{1}{Upper Bound}
	We use the notations $D_q(n,k), d_q(n,k)$ from the proof of Lemma~\ref{lem:bounds} and consider the set $G(m) = D_q(n,k)^m$. We also denote $g(m) = |G(m)| = d_q(n,k)^m$. Note that the set $G(m)$ satisfies the $k$-RLL constraint, and hence 
	$$\lim_{m\rightarrow \infty} \frac{\log(g(m))}{nm}\leq E_{k,q}.$$
	We therefore get 
	$$\lim_{m\rightarrow \infty} \frac{\log(d_q(n,k)^m)}{nm} = \frac{\log(d_q(n,k))}{n}  \leq E_{k,q}.$$
	By using Equation~(\ref{eq:d_nk}) we have
	$$\frac{\log(a_q(n,k) - 2q^{n-\lceil{k/2}\rceil})}{n}  \leq E_{k,q},$$
	and 
	$${a_q(n,k)} \leq 2^{nE_{k,q}} + 2q^{n-\lceil{k/2}\rceil}.$$
	
	\proofpart{2}{Lower Bound}
	For all positive integer $m$ we have 	$A_q(mn,k)\subseteq A_q(n,k)^m$ and therefore we deduce that 
	\begin{align*}
	E_{k,q}&= \lim_{m\rightarrow\infty}\frac{\log(a_q(mn,k))}{mn}\\ 
	&\le \lim_{m\rightarrow\infty}\frac{\log(a_q(n,k)^m)}{mn} \\
	&= \frac{\log(a_q(n,k))}{n},
	\end{align*}
	and the lower bound follows directly.
\end{IEEEproof}

For some values of $k$ the result from Lemma~\ref{lem:cardinality} gives weaker bounds than the one presented in Lemma~\ref{lem:bounds}. However, for the case of $k=\lceil \log_q n\rceil + z, z\in \mathbb{Z}$, we next show that the lower and upper bound of Lemma~\ref{lem:cardinality} are asymptotically tight. 

Recall that Jain et. al showed in \cite{J17} that 
\begin{equation} 
	E_{k,q}=\log q - \frac{(q-1) \log e}{ q^{k+2}}(1 + o(1))
\end{equation}
or in other words,
\begin{equation} \label{eq:Jain1}
\lim_{k\rightarrow \infty} \frac{(q-1)\log e q^{-k -2}}{\log q- E_{k,q}} =1.
\end{equation}

According to this result and the properties proved in Proposition~\ref{pro:powapprox} and Lemma~\ref{lem:cap} we conclude in Lemma~\ref{lem:capcity} what the asymptotic behavior of $2^{nE_{k,q}}$ is. Then, in Theorem~\ref{the:main} we show that $2^{n-\lceil{k/2}\rceil+1}$ is negligible relatively to $2^{nE_{k,q}}$, therefore the bounds in Lemma~\ref{lem:cardinality} meet and the asymptotic behavior of $a_q(n,k)$ is established.      

The following proposition will be in use in the proof of Lemma~\ref{lem:capcity} and its proof is given in Appendix~\ref{app:sec3}.   
\begin{proposition} \label{pro:powapprox}
	Let $f(n),\ g(n)$ be functions such that $\lim_{n\rightarrow \infty} g(n) = 1$ and $1\le f(n) \le C$ for a constants $C$. Then,
	$$ f(n)^{g(n)}\approx f(n).$$
\end{proposition}

Note that the requirement that $f(n)$ is bounded is essential, otherwise the proposition does not hold. For example if $f(n)=2^n$ and $g(n)=1+\frac{1}{n}$ we have that 	
$\lim_{n\rightarrow \infty} \frac{f(n)^{g(n)}}{f(n)}=2.$
\begin{lemma} \label{lem:cap}
	There exists an integer $N$ such that for all $n\ge N $ and $k=\lceil \log_q n\rceil + z, z\in \mathbb{Z}$ 
	$$\log q -\frac{C}{n} \le E_{k,q} \le \log q, $$ 
	for some constant $C>0$ which is independent of $n$.
\end{lemma}

\begin{IEEEproof}
	First, $ E_{k,q} \le \log q$ from the definition of $E_{k,q}$.
	From Corollary~\ref{cor:theta} there exists a constant $C'$ such that for large enough $n$ 
	$$\textmd{red}(A_q(n,\lceil \log_q n\rceil + z))\le C',$$ and equivalently $a_q(n,\lceil \log_q n\rceil + z) \ge q^{n-C'}$. From Lemma~\ref{lem:cardinality}
	\begin{align*}
	E_{k,q} &\ge \frac{\log(a_q(n,k) - 2q^{n-\lceil{k/2}\rceil})}{n}\\
	&\ge  \frac{\log(q^{n-C' } - 2q^{n-\lceil{k/2}\rceil})}{n} \\
	&=  \frac{n \log q + \log(q^{-C'} - 2q^{-\lceil{k/2}\rceil})}{n}. 
	\end{align*}
	There exists an integer $N$ such that for all $n\ge N,\  2q^{-\lceil{k/2}\rceil}\le 0.5q^{-C'}$. Thus,
	\begin{align*}
	E_{k,q} &\ge \frac{n \log q+ \log(q^{-C'} - 0.5q^{-C'})}{n}\\
	&=\frac{n \log q+ \log(0.5q^{-C'})}{n}\\
	&=\log q-\frac{1+C'\log q}{n}
	\end{align*}
	and by choosing $C=1+C'\log q$ the result follows. 
\end{IEEEproof}

We are now ready to use the result of Jain et al. mentioned in Equation~(\ref{Jain}), in order to establish the following result.
\begin{lemma} \label{lem:capcity}
	For $k=\lceil \log _qn\rceil + z, z\in \mathbb{Z}$,
	$$2^{nE_{k,q}}\approx \frac{q^n}{e^{(q-1) q^{\Delta_n - z -1 }}} ,$$ 
	where $\Delta_n = \log_q n -\lceil \log_q n\rceil$.
\end{lemma}

\begin{IEEEproof}
	We use Proposition~\ref{pro:powapprox} with $f(n)= 2^{n(\log q-E_{k,q})}$ and $g(n)= \frac{(q-1)\log e q^{-k -2}}{\log q- E_{k,q}}$.
	From Lemma \ref{lem:cap}, when $k=\lceil \log_q n\rceil + z,$  
	$$1\le2^{n(\log q-E_{k,q})}=f(n) \le 2^C. $$
	From Eq. (\ref{eq:Jain1}) $\lim_{n\rightarrow \infty} g(n) = 1$ thus the requirements of the proposition are satisfied and we get  
	$$2^{n(\log q-E_{k,q})}\approx2^{n(q-1)\log e q^{-(k-1) -2}}.$$
	We conclude that
	\begin{align*}
	2^{nE_{k,q}} &= 2^{n\log q+n(E_{k,q}-\log q)}\\
	&\approx 2^{n\log q-n(q-1)\log e q^{-k-1}} \\
	&\approx 2^{n\log q-n(q-1)\log e q^{-\lceil \log_q n\rceil - z-1}} \\
	&\approx \frac{q^n}{e^{ (q-1)q^{\Delta_n - z -1 }}},
	\end{align*}
	where $\Delta_n = \log_q n -\lceil \log_q n\rceil$.
\end{IEEEproof}

Finally, we can now apply the result of Lemma~\ref{lem:capcity} in Lemma~\ref{lem:cardinality} and obtain the asymptotic behavior of $a_q(n,\lceil \log_q n\rceil + z)$ in the following theorem.
\begin{theorem} \label{the:main} 
	For $k=\lceil \log_q n\rceil + z$, $z\in \mathbb{Z}$,
	$$a_q(n,k)\approx \frac{q^n}{e^{(q-1)q^{\Delta_n - z -1 }}},$$ 
	where $\Delta_n = \log n -\lceil \log n\rceil$.
\end{theorem}

\begin{IEEEproof}
	Lemma~\ref{lem:cardinality} gives us
	$$1\leq \frac {a_q(n,k)} {2^{nE_{k,q}}} \leq 1 + \frac{2q^{n-\lceil{k/2}\rceil}}{2^{nE_{k,q}}}.$$
	By using Lemma~\ref{lem:capcity} we get that for $k=\lceil \log _q n\rceil + z$
	$$\lim_{n\rightarrow \infty} \frac{2q^{n-\lceil{k/2}\rceil}}{2^{nE_{k,q}}} = \lim_{n\rightarrow \infty} \frac{2q^{n-\lceil{k/2}\rceil}}{{q^n}{e^{ -(q-1)q^{\Delta_n - z -1 }}}} =0,$$ 
	and conclude that 
	$$1\leq \lim_{n\rightarrow \infty} \frac {a_q(n,k)} {2^{nE_{k,q}}} \leq \lim_{n\rightarrow \infty} 1 + \frac{2q^{n-\lceil{k/2}\rceil}}{2^{nE_{k,q}}} = 1.$$
	Therefore, $$a_q(n,k)\approx2^{nE_{k,q}},$$ and together with Lemma~\ref{lem:capcity} the result follows directly. 
	
\end{IEEEproof}

\begin{remark}
	We would like to note that the results in this section match the results by Schilling~\cite{S12} on the distribution of the longest runs in arbitrary vectors. In particular, it is stated in~\cite{S12} that the typical length of the longest run in $n$ flips of a fair coin converges to $\log n-1$. However, we find the proof in this section to be more accurate for the purpose of exactly calculating the number of redundancy bits of $k$-RLL codes.
\end{remark}

\subsection{Efficient Encoding and Decoding Algorithm} \label{subsec:encoding}
In this subsection we provide an algorithm to efficiently encode and decode vectors which avoid zeros run of length $\lceil\log_q n\rceil +1$. We show the algorithm for the binary case, however, it is straightforward to extend it for $q>2$. According to Theorem~\ref{the:main} the redundancy of $A_2(n, \log n +1)$ is approximately $\frac{\log e}{4}\approx0.36$ when $n=2^i, i\in \mathbb{N}$. The algorithm described next uses one redundancy bit, however, it has linear encoding and decoding complexities. 
\begin{algorithm}\begin{small}	
		\caption{Zero Run-Length Encoding}\label{alg:encoding}
		\begin{algorithmic}[1]
			\vspace{.1ex}
			\Require Sequence $\vec{x}\in \Sigma^{n'},\ n'\le n $ 
			\Ensure $\vec{y} \in \Sigma^{n'+1}$ with zeros run length $\leq \lceil\log n\rceil$
			\State Define $\vec{y} = \vec{x}1 \in \Sigma^{n'+1}$
			\State Set $i=1$ and $i_{end} = n'$
			\While{$i \leq i_{end}-\lceil\log n\rceil$}
			\If{ $w_H(\vec{y}_i^{i+\lceil\log n\rceil})=0$} 
			\State remove the zeros run $\vec{y}_i^{i+\lceil\log n\rceil}$ from $\vec{y}$
			\State $\vec{p(i)}$: binary representation of $i$ with $\lceil\log n\rceil$ bits
			\State append $\vec{p(i)}0$ to the right of $\vec{y}$
			\State set $i_{end}=i_{end}-\lceil\log n\rceil-1$
			\Else
			\State set $i=i+1$
			\EndIf
			\EndWhile
		\end{algorithmic}\end{small}
	\end{algorithm}
	
	The following lemma proves the correctness of Algorithm~\ref{alg:encoding}.
	\begin{lemma}\label{lem:encoding}
		For all $n'\leq n$, given any vector $\vec{x} \in \Sigma^{n'}$, Algorithm~\ref{alg:encoding} outputs a sequence $\vec{y}\in \Sigma^{n'+1}$, where any zeros run has length at most $\lceil\log n\rceil$ and such that $\vec{x}$ can be uniquely reconstructed given $\vec{y}$. Furthermore, the time and space complexity of the algorithm and its inverse is $\Theta(n)$.
	\end{lemma}
	\begin{IEEEproof}
		The algorithm starts by initializing $\vec{y}=\vec{x}1$. We then iterate over the indices of $\vec{y}$ that correspond to the indices of the input word $\vec{x}$. If we encounter an index in which a $\lceil\log n\rceil +1 $ zeros run starts, we remove the run and append $\vec{p(i)}0$, which we call a pointer, to the right of $\vec{y}$, where $\vec{p(i)}$ is the binary representation of the index $i$. 
		
		First, notice that each appended pointer $\vec{p(i)}0$ has the same length as the corresponding removed run. Therefore throughout the algorithm the length of $\vec{y}$ does not change. There exists an index $1\le t \le n'$ such that the output $\vec{y}$ is of the form $\vec{y}_1^t1\vec{y}_{t+2}^{n'+1}$ where $\vec{y}_1^t$ is the remainder of $\vec{x}$ after removing the zeros runs and $\vec{y}_{t+2}^{n'+1}$ is the list of the pointers $(\vec{p(i)},0)$ representing the indices of the removed zeros runs. 
		
		To reconstruct $\vec{x}$ given $\vec{y}$ we start by locating the separating bit 1 on position $t$. We start from the right and check whether the rightmost bit is 1 or 0. In case it is 0, the $\lceil\log n\rceil$ bits to the left correspond to a pointer, we skip them and repeat the process until we encounter the separating 1. We then construct the original $\vec{x}$ by inserting zeros runs of length $\lceil\log n\rceil+1$ to the remainder part $\vec{y}_1^t$ according to the pointers part $\vec{y}_{t+2}^{n'+1}$. 
		
		Next, we show that $\vec{y}$ does not contain a zeros run of length greater than $\lceil\log n\rceil$. It is clear that $\vec{y}_1^t$ does not contain such a run. The separating 1 ensures that there is no zeros run which starts in $\vec{y}_1^t$ and ends in $\vec{y}_{t+2}^{n'+1}$. The structure of $\vec{y}_{t+2}^{n'+1}$ is a sequence of concatenated pointers of the form $\vec{p(i)}0$. It suffices to show that any sub-vector of $\vec{y}_{t+2}^{n'+1}$ of the form $\vec{p(j)}0\vec{p(k)}$ does not consist a zeros run of length greater than $\lceil\log n\rceil$. Here, $j$ and $k$ represent the indices of two consecutive locations where a $\lceil\log n\rceil +1$ zeros run was found in the while loop in the algorithm and note that $0<j\le k$. 
		
		We consider the leftmost one in the binary representations $\vec{p(j)}$ and $\vec{p(k)}$. Since $j \le k$, the position of the leftmost one within $\vec{p(j)}$ is smaller or equal to the position of the leftmost one within $\vec{p(k)}$, thus any window of length $\lceil\log n\rceil + 1$ must contain at least one of the leftmost ones from $\vec{p(j)}$ and $\vec{p(k)}$. 
		
		Note that since the indices of the input vector $\vec{x}$ are indexed starting from 1, we do not write 0 in any of the pointers. We do not write $n'$ either since the largest index we consider for zeros runs is $n'-\lceil \log n\rceil$. Lastly, the complexity of the algorithm is $\cO(n)$ since the complexity of each pointer update when encountering a zeros run is $\Theta(\log n)$ and the number of these operations is at most $n'/\log n$.
	\end{IEEEproof}
	
	The following example demonstrates how the encoding in Algorithm~\ref{alg:encoding} works.
	\begin{example}
		Let $n'=n=13$ and therefore $\lceil\log n\rceil = 4$ and $\lceil\log n\rceil+1 = 5$. Consider the following sequence:
		\begin{equation*}
		\vec{x} = 1000 0000 0000 1,
		\end{equation*}
		Let us go through the steps of Algorithm~\ref{alg:encoding}.
		\begin{enumerate}
			\item $\vec{y} = \vec{x}1 = 1000 0000 0000 11$
			\item $i=1$ and $i_{end} = 13$
			\item $i=1$: $w_H(\vec{y}_1^5)\ne0, \ i=i+1$ 
			\item $i=2$: $w_H(\vec{y}_2^6)=0$,
			\begin{enumerate}
				\item Remove $\vec{y}_2^6$ from $\vec{y}: \ \vec{y} = 100 0000 11$
				\item Define $\vec{p(2)} = 0010$
				\item Append $\vec{p(2)}0 = 00100: \ \vec{y} = 100 0000 1100100$
				\item Set $i_{end} = 13-5=8$
			\end{enumerate}
			\item $i=2$: $w_H(\vec{y}_2^6)=0$,
			\begin{enumerate}
				\item Remove $\vec{y}_2^6$ from $\vec{y}: \ \vec{y} = 10 1100100$
				\item Define $\vec{p(2)} = 0010$
				\item Append $\vec{p(2)}0 = 00100: \ \vec{y} = 10 110010000100$
				\item Set $i_{end} = 8-5=3$
			\end{enumerate}
			\item $i=2$: $w_H(\vec{y}_2^6)\ne0, \ i=i+1$.
		\end{enumerate}
		The decoding works as described in the proof of Lemma~\ref{lem:encoding}.
	\end{example}
	
	A similar algorithm to the problem solved by Algorithm~\ref{alg:encoding} was recently proposed in~\cite{S16} to efficiently encode sequences that do not contain runs of zeros and ones of length $k$ with a single redundancy bit.
	Specifically, in~\cite{S16} the authors showed how to accomplish this task with $k=\lceil\log n\rceil+4$. Algorithm~\ref{alg:encoding} can be slightly adjusted in order to solve the problem in~\cite{S16} with $k=\lceil\log n\rceil+2$. 
	Lastly, we note that Kauts presented in~\cite{K65} an algorithm which encodes all words avoiding zeros runs of any specific length with optimal redundancy. However, the space complexity of this algorithm is $\Theta(n^2)$.
	
	Algorithm~\ref{alg:encoding} solves the problem of avoiding zeros runs of length $\lceil \log n \rceil +1$. However, it can be extended to avoid zeros runs of any length in the following manner. Assume for example that we want to avoid zeros runs of length $0.5 \log n$. We start with a length-$n$ vector, divide it into blocks of length $\frac{\sqrt{n}}{2}$ and apply Algorithm~\ref{alg:encoding} on each block since $0.5 \log n = \log(\frac{\sqrt{n}}{2})+1$. We append 1 to each of the output vectors and the final resulting vector is the concatenation of all of them. This approach can be applied for any value of $k$ and it achieves optimal order of redundancy, with linear encoding and decoding complexities.   
	   
\section{Mutually Uncorrelated Codes}\label{sec:MUCodes}
	In this section we expand the study of MU codes. Specifically, we show that the lower bounds in~(\ref{eq:mulb1}),~(\ref{eq:mulb2}) are asymptotically tight.
	
	We are interested in maximizing the value of $|\cC_1(n,q,k)|$ over all values of $k$. Notice that $$|\cC_1(n,q,k)|=(q-1)^2a_q(n-k-2,k),$$ and therefore the proof of the main theorem in this section highly relies on the analysis in Section~\ref{sec:RLL}. However, in Section~\ref{sec:RLL} we analyzed the asymptotic size $a_q(n,k)$ while we are actually interested in $a_q(n-k-2,k)$. We chose to present the analysis of $a_q(n,k)$ in Section~\ref{sec:RLL} since it is similar to  the analysis of $a_q(n-k-2,k)$ and we believe it will be of use in other problems as well. In this section we complete some missing parts to obtain the asymptotic behavior of the maximal value of $|\cC_1(n,q,k)|=(q-1)^2a_q(n-k-2,k)$. We maily focus on the trade-off formed by the choice of $k$: on one hand, reducing the value of $k$ results with a larger value for $n-k-2$ and thus smaller redundancy due to the fixed $k$-length prefix in the construction. On the other hand, smaller $k$ implies a stronger $k$-RLL constraint which requires larger redundancy in the remaining part.
	
	We start by showing that it is sufficient to look only into values of $k$ which are of the form $\lceil\log_q n\rceil +z, z\in \mathbb{Z}$. 
	\begin{lemma} 
		The value of $k$ that minimizes the redundancy of $\cC_1(n,q,k)$ is $\lceil\log_q n\rceil +z, z\in \mathbb{Z}$. 
	\end{lemma}
	
	\begin{IEEEproof}From Corollary~\ref{cor:theta}, in the case of $k=\lceil\log_q n\rceil +z, z\in \mathbb{Z}$ the redundancy of $\cC_1(n,q,k)$ is $k+\Theta(1)=\lceil\log_q n\rceil+z+\Theta(1)$. Since the redundancy of $\cC_1(n,q,k)$ is at least $k$, greater values of $k$ i.e., $k=\lceil\log_q n\rceil+\small{\omega}(1)$, lead to higher redundancy, and so we disregard them. For values $k$ of the form $\log_q n-f(n)$, where $f(n)=\small{\omega}(1)$ we again turn to Corollary~\ref{cor:theta} to claim that the redundancy is $\Theta(k+2^{f(n)})= \Theta(\log_q n-f(n)+2^{f(n)})$ which is also asymptoticly greater than $\lceil\log_q n\rceil+z+\Theta(1)$. We therefore summarize that the minimal redundancy, or the maximal cardinality of $\cC_1(n,q,k)$ is achieved when $k$ is of the form $k=\lceil\log_q n\rceil +z, z\in \mathbb{Z}$.
	\end{IEEEproof}	
		
	We next look further into the specific choice of $k$ that maximizes the cardinality and obtain the asymptotic behavior of the maximal cardinality.
	
	For the rest of this section we denote $n'=n-k-2$. From Lemma~\ref{lem:cardinality} we have that 
	\begin{equation} 
	2^{n'E_{k,q}}\leq {a_q(n',k)} \leq 2^{n'E_{k,q}} + 2q^{n'-\lceil{k/2}\rceil}. \label{eq:mucardinality}
	\end{equation}
	In the following lemma we establish how $2^{n'E_{k,q}}$ behaves asymptotically for the values of $k$ of interest to us. The proofs of Lemma~\ref{lem:mucapacity} and Lemma~\ref{lem:mumain} are attached in Appendix~\ref{app:sec4} as they share similar ideas to the proofs in Section~\ref{sec:RLL}. 
	\begin{lemma} \label{lem:mucapacity}
		For $k=\lceil \log_q n\rceil + z, z\in \mathbb{Z}$ ,
		$$2^{n'E_{k,q}}\approx \frac{q^n}{n} \cdot \frac{q^{\Delta_n-z-2}}{e^{(q-1) q^{\Delta_n - z-1}}} ,$$ 
		where $\Delta_n = \log_q n -\lceil \log_q n\rceil$.
	\end{lemma}

We apply the result of Lemma~\ref{lem:mucapacity} in the inequality (\ref{eq:mucardinality}) to obtain the asymptotic behavior of $a_q(n',\lceil \log n\rceil + z)$. 
	\begin{lemma} \label{lem:mumain} 
	For $z\in \mathbb{Z}$, 
	$$|\cC_1(n,q,\lceil \log n\rceil + z)|\approx \frac{q^n}{n}\big(\frac{q-1}{q}\big)^2q^{\Delta_n-z-\log_q e (q-1) q^{\Delta_n - z-1 }}$$
	where $\Delta_n = \log_q n -\lceil \log_q n\rceil$.
	\end{lemma}
	
Next we optimize this term over all values of $z\in \mathbb{Z}$ in order to establish the maximal cardinality of $\cC_1(n,q,k)$.
	\begin{theorem}\label{th:c1size} 
	
	\[\cC_1(n,q) \approx \frac{q^n}{n}\cdot \big(\frac{q-1}{q}\big)^2 q^{F(\Delta_n)}\le  \frac{q^n}{n}\cdot\frac{q-1}{eq},\]
	where $\Delta_n= \log_q n-\lceil\log_q n\rceil$ and $$F(\Delta_n) = \max_{z\in \{-2,-1,0\}} \big\{\Delta_n-z - \log_q(e)(q-1)q^{\Delta_n-z-1}\big\}. $$ The inequality is tight when $n\rightarrow\infty$ over any subsequence of $n$ that satisfies $\Delta_n=-\log_q(q-1)$.

	\end{theorem}
	\begin{IEEEproof}
		From Lemma~\ref{lem:mumain}, when $k=\lceil \log_q n\rceil + z$ we get
		\begin{align*}		
		|\cC_1(n,q,k)| \approx \frac{q^n}{n}\big(\frac{q-1}{q}\big)^2q^{\Delta_n-z-\log_q e (q-1) q^{\Delta_n - z-1 }}.
		\end{align*}
		Let us denote 
		$$f(\Delta_n,z)=\Delta_n-z-\log_q e (q-1) q^{\Delta_n - z-1 }.$$ 
		We are interested in the value of $z\in \mathbb{Z}$ that maximizes the size of $\cC_1(n,q,k)$, that is, the value of $z\in \mathbb{Z}$ that maximizes $f(\Delta_n,z)$ for each $\Delta_n$. 
		$$\frac{\partial{f} }{\partial {z}} = -1+(q-1)q^{\Delta_n - z-1},$$
		so the only maximum of the function $f(\Delta_n,z)$ is achieved for $z_0=\log_q(q-1)-1+\Delta_n$. However, $z_0$ is not necessarily an integer while we are interested in integers only. Since $-1<\Delta_n\le 0$ we have $-2 < z_0 < 0$ thus the maximum over $z\in \mathbb{Z}$ is achieved by one of the options $z\in \{-2,-1,0\}$. 
		
		We therefore obtain the following:
		\[\cC_1(n,q) \approx \frac{q^n}{n}\big(\frac{q-1}{q}\big)^2 q^{F(\Delta_n)}, \]
		where $\Delta_n= \log_q n-\lceil\log_q n\rceil$ and $$F(\Delta_n) = \max_{z\in \{-2,-1,0\}} \big\{\Delta_n-z - \log_q(e)(q-1)q^{\Delta_n-z-1}\big\}. $$\
		In Appendix~\ref{app:sec4} we analyze how $F(\Delta_n)$ behaves for each value of $\Delta_n$ and obtain that when $q=2$
		$$F(\Delta_n)=
		\begin{cases}
		f(\Delta_n, -2),\  for\  -1<\Delta_n\le \log (\ln 2) \\
		f(\Delta_n, -1),\   otherwise  \\
		\end{cases}
		$$ 
		and when $q>2$
		$$F(\Delta_n)=
		\begin{cases}
		f(\Delta_n, -1),\  for\  -1<\Delta_n\le \delta_0 \\
		f(\Delta_n, 0),\  \ \ otherwise  \\
		\end{cases}
		$$  
		for $\delta_0=-\log_q\frac{(q-1)^2}{q\ln q}$.
	   We also discuss in Appendix~\ref{app:sec4} the maximal value of $F(\Delta_n)$ which leads to the maximal cardinality $\frac{q^n}{n}\cdot\frac{q-1}{eq}$ 
\end{IEEEproof}

The result of Theorem~\ref{th:c1size} aligns with the results from~\cite{C13,G60,L64} which we recalled in Equations (\ref{eq:mulb1}) and (\ref{eq:mulb2}). The lower bound from (\ref{eq:mulb1}) states that 
\begin{equation}
\cC_1(n,q) \gtrsim q^{-\frac{q}{q-1}}\ln q \frac{q^n}{n} \tag{\ref{eq:mulb1}}
\end{equation}
 and it holds when $n\rightarrow \infty$ over any series of $n$. The lower bound (\ref{eq:mulb2}) claims that 
 \begin{equation}
	\cC_1(n,q) \gtrsim \frac{q-1}{qe}\cdot \frac{q^n}{n}\tag{\ref{eq:mulb2}}
 \end{equation}
 and it holds when $n\rightarrow \infty$ over the subseries $\frac{q^i-1}{q-1}, i\in \mathbb{N}$. Observe that when $n=\frac{q^i-1}{q-1},\ \lim_{n\rightarrow \infty} \Delta_n = -\log_q(q-1)$ thus the result of Theorem~\ref{th:c1size} and the bound (\ref{eq:mulb2}) agree.

\begin{figure}
	\begin{center}
		\begin{tabular}{@{}c@{}}
			\includegraphics[width =1.1\linewidth]{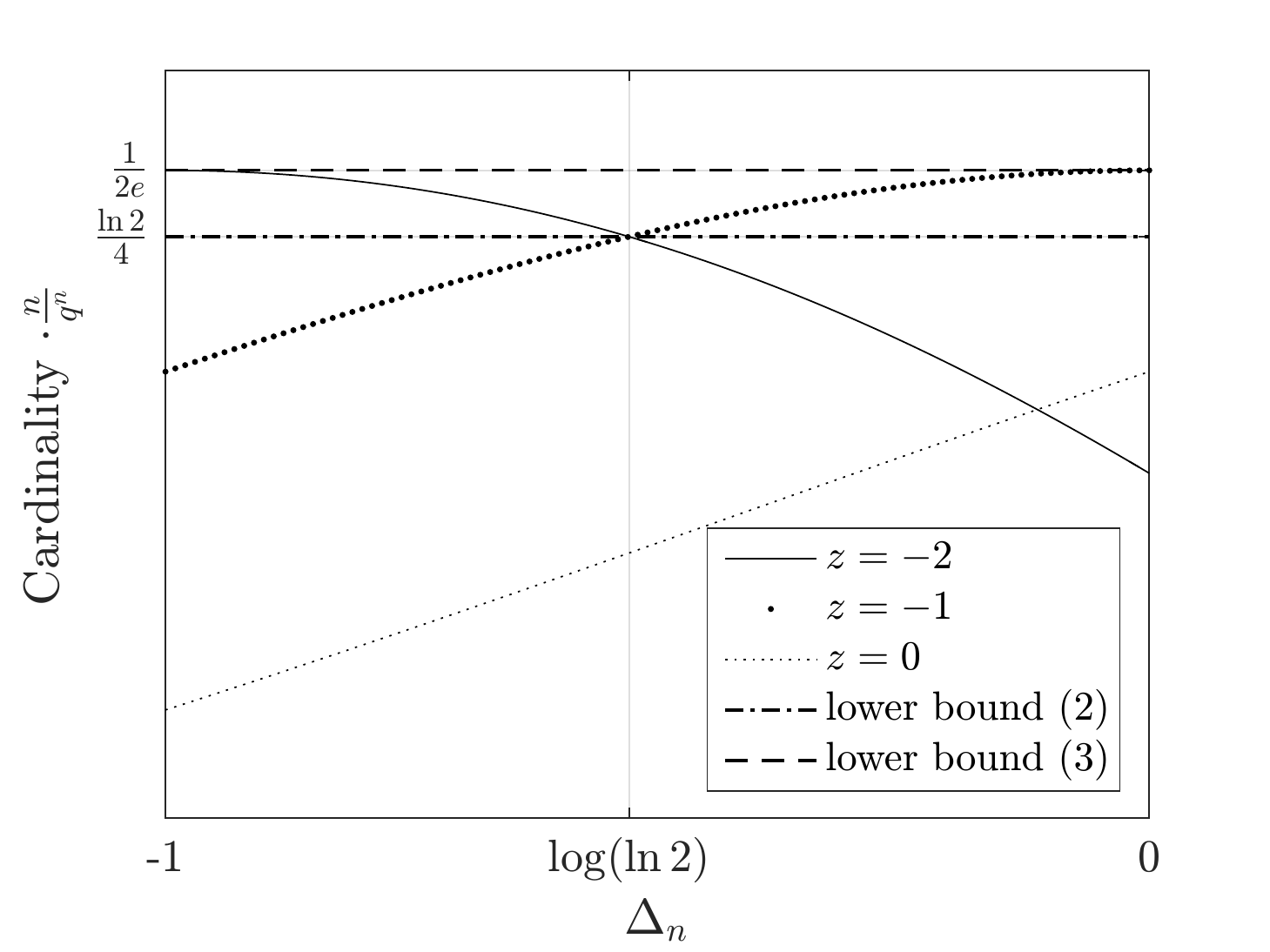} \\
			\small (a) Cardinalities for $q=2$ \label{fig:q2}
		\end{tabular}
		\begin{tabular}{@{}c@{}}
			\includegraphics[width =1.1\linewidth]{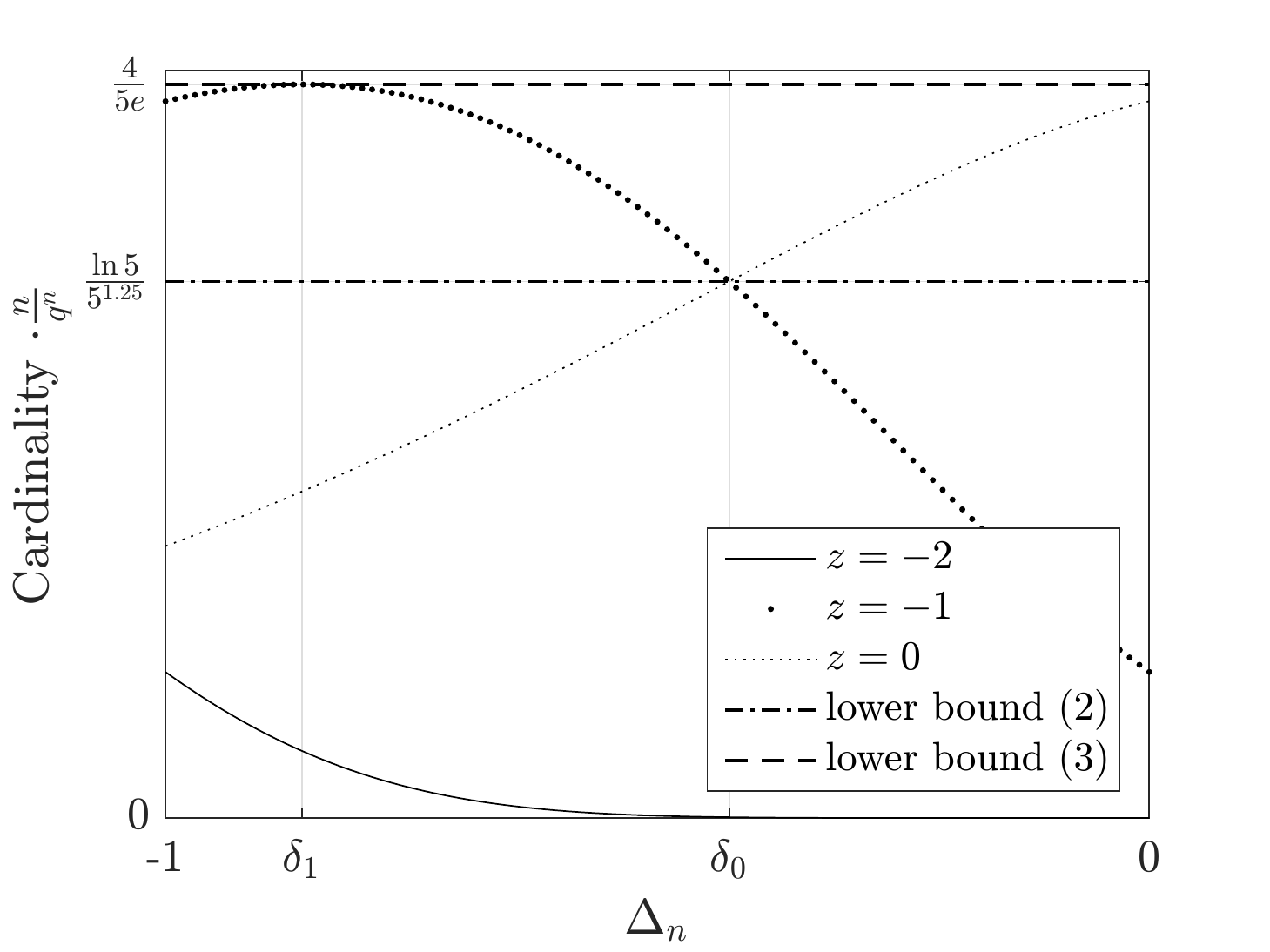} \\
			\small (b) Cardinalities for $q=5$ \label{fig:q5}
		\end{tabular}
	\end{center}
	\caption{Comparison between the construction's cardinality according to Lemma~\ref{lem:mumain}, multiplied by $\frac{n}{q^n}$ for different values of $z$, and the bounds (\ref{eq:mulb1}) and (\ref{eq:mulb2}) from~\cite{C13,G60,L64}} \label{fig:q25}.
\end{figure}

\begin{figure}
\begin{center}
	\includegraphics[width =1.1\linewidth]{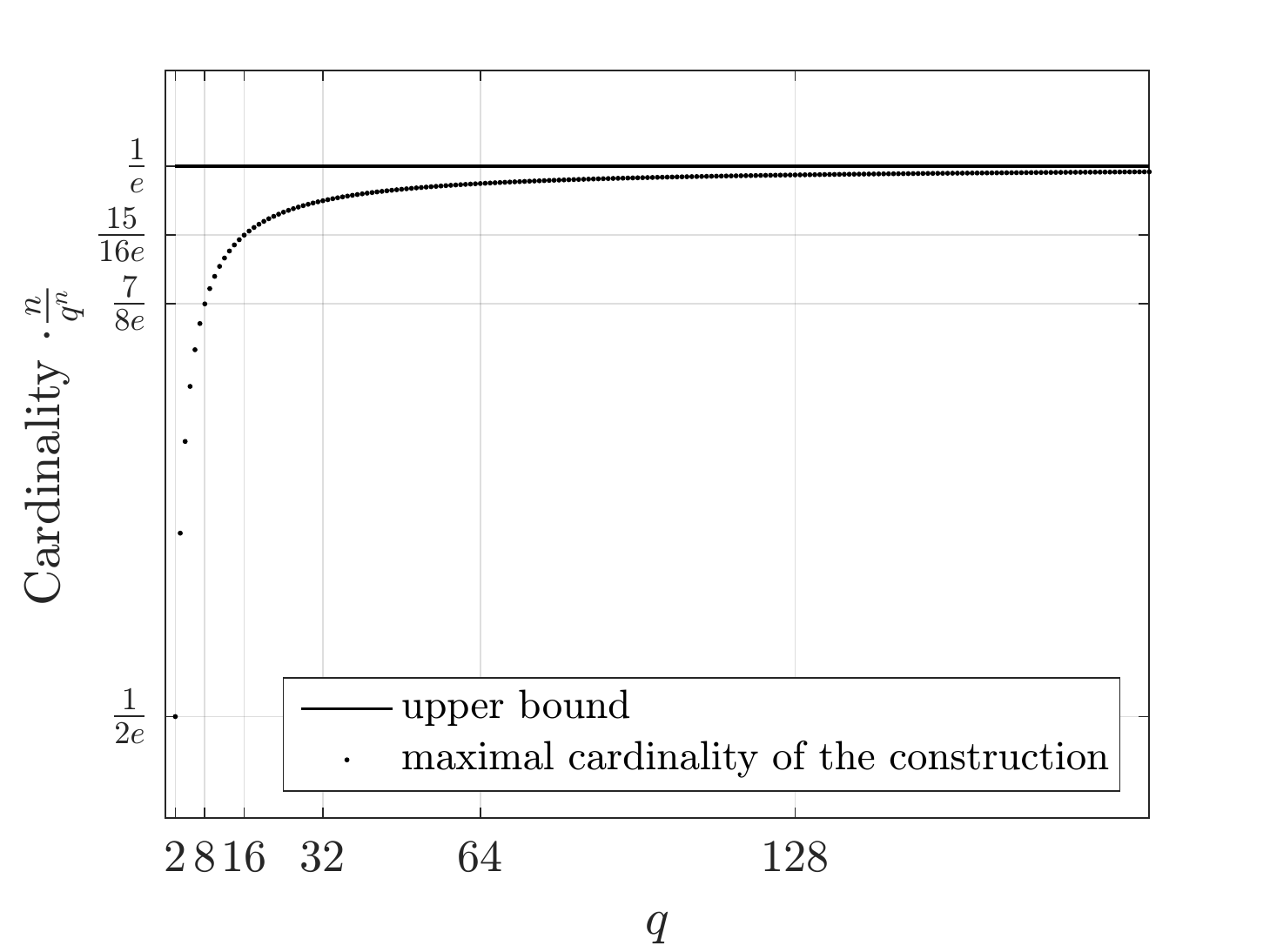} \caption{A comparison between the construction's maximal cardinality according to Theorem~\ref{th:c1size} and the upper bound of the cardinality of MU codes from~\cite{L70}, mentioned in (\ref{eq:ub}).} \label{fig:qinf}
\end{center}
\end{figure}

The bounds in (\ref{eq:mulb1}) and (\ref{eq:mulb2}) provide lower asymptotic bounds on the cardinality $\cC_1(n,q)$, while in Theorem~\ref{th:c1size} we obtain the explicit expression of the asymptotic behavior for any value of $\Delta_n$ and show that the lower bounds are asymptotically tight. According to Lemma~\ref{lem:mumain} the term $$|\cC_1(n,q,\lceil \log_q n\rceil + z)| \cdot \frac{n}{q^n}$$ depends only on $z,\ q$ and $\Delta_n = \log_q n -\lceil \log_q n\rceil$. Specifically it does not grow with $n$. In Fig.~\ref{fig:q25} we plot this term for the three interesting values of $z \in \{-2,-1,0\}$ as a function of $\Delta_n$, alongside the lower bounds from (\ref{eq:mulb1}) and (\ref{eq:mulb2}). The first, second graph in Fig.~\ref{fig:q25} corresponds to $q=2$, $q=5$, respectively. When $q=5$ for example, the maximal cardinality is achieved when $z=-2$ or when $z=-1$, depending on the value of $\Delta_n$. Among the range of values of $\Delta_n$, $\delta_0=-\log_5(\frac{16}{5ln5})\approx -0.43$ yields the minimal cardinality which aligns with the bound (\ref{eq:mulb1}). The maximal cardinality is achieved for $\delta_1=-\log_q(q-1)=-\log_5 4 \approx -0.86$ where it meets the bound (\ref{eq:mulb2}), and the maximal cardinality is $\frac{q-1}{qe}\cdot \frac{q^n}{n}$.
So far we referred to $q$ as a constant, however, an interesting fact to notice when $q\rightarrow\infty$ and $n\rightarrow\infty$, the maximal cardinality $\frac{q-1}{qe}\cdot \frac{q^n}{n}$ approaches the upper bound from~\cite{L70}, $\frac{q^n}{e(n-1)}$. This is illustrated in Fig.~\ref{fig:qinf}. Having said that, note that in the binary case there is a gap of factor 2 between the maximal cardinality and the upper bound and closing this gap remains an open problem.


\section{The Window Weight Limited Constraint}\label{sec:WWL}

In this section we introduce a natural extension to the RLL constraint which we call the \emph{window weight limited} constraint. We study this constraint for the purpose of constructing a new family of codes, called \emph{$(d_h,d_m)$-Mutually Uncorrelated Codes}, which will be presented later in Section~\ref{sec:ECC_MUCodes}. Our main results in this section are an upper bound on the size of the set of vectors satisfying the window weight limited constraint and a construction with efficient encoding and decoding, and almost optimal cardinality. We start with the definition of this constraint.
\begin{definition}
	Let $d$ and $k$ be positive integers. We say that a vector $\vec{a}\in \Sigma_q^n$ satisfies the $(\mathit{d},k)$\textbf{-window weight limited (WWL)} constraint, and is called a $(d,k)$\textbf{-WWL} vector, if $n<k$ or $\forall i\in [n-k+1]: w_H(\vec{a}_i^{i+k-1})\ge d$.
\end{definition}
We call a set of $(d,k)$-WWL vectors a \emph{$(d,k)$-WWL code} and we denote by $A_q(n,k,d)$ the set of all $(d,k)$-WWL vectors over $\Sigma_q^n$. Lastly, $a_q(n,k,d) = |A_q(n,k,d)|$. 

This constraint states that a vector $\vec{a}\in\Sigma_q^n$ is a $(d,k)$-WWL vector if the Hamming weight of every consecutive subsequence of length $k$ in $\vec{a}$ is at least $d$. A similar constraint was studied in~\cite{Q13} when studying time space codes for phase change memories. However, the authors of~\cite{Q13} studied the opposite constraint in which the weight of every window of length $k$ is \emph{at most} $d$. Furthermore, as will be explained later, we are interested in the case where $k$ depends on the word length $n$ as opposed to the analysis in~\cite{Q13} where $k$ is fixed. 
Notice that the $(1,k)$-WWL constraint is equivalent to the $k$-RLL constraint.
In the next lemma we provide an upper bound on the size of $A_q(n,k,d)$. The proof is deferred to Appendix~\ref{app:sec5} since it shares similar ideas with the proof of the upper bound in Lemma~\ref{lem:bounds}.

\begin{lemma} \label{lem:bounds2} 
	Let $n,k,d$ be positive integers such that $d\le k\le n$. Then, there exists a constant $C>0$ such that for $n$ large enough
	$$a_q(n,k,d) \le q^{n-C\frac{(n-2k)k^{d-1}}{q^k}}.$$ 
\end{lemma}

For the rest of the paper we let $\mathcal{F}(n,d)$ be 
\begin{equation}\label{eq:Fnd}
\mathcal{F}(n,d)= \lceil\log_q n \rceil + (d-1) (\lceil\log_q \lceil \log_q n \rceil \rceil + C ) + 2,
\end{equation} 
where $C$ is a constant equal to the minimum integer such that $\lceil\log_q \lceil \log_q n \rceil \rceil + C \ge \lceil\log_q(\mathcal{F}(n,d)+2)\rceil$. 
From Lemma~\ref{lem:bounds2} we also have that 
\begin{align*}
C\frac{(n-2k)k^{d-1}}{q^k} \le  \textrm{red}(A_q(n,k,d)),
\end{align*}
where $C>0$ is a constant. By setting $k$ to be $\mathcal{F}(n,d)-f(n), \ f(n)>0$  we get that the redundancy of $A_q(n,\mathcal{F}(n,d)-f(n),d)$ is $\Omega(q^{f(n)}).$

Next we present an explicit algorithm for encoding and decoding $(d,\mathcal{F}(n,d))$-WWL vectors with $n'\le n$ information bits and $d$ redundancy symbols. 
\begin{algorithm}\begin{small}
		\caption{Window Weight Limited Encoding}\label{alg:wwl-encoding}
		\begin{algorithmic}[1]
			\vspace{.1ex}
			\Require $\vec{x}\in \Sigma_q^{n'}$ and an integer $d>1$
			\Ensure $(d,\mathcal{F}(n,d)) $-WWL vector $\vec{y} \in \Sigma_q^{n'+d}$ 
			\State Define $\vec{y} = \vec{x}1^{d} \in \Sigma_q^{n'+d}$
			\State Set $i=1$ and $i_{end} = n'$
			\While{$i \leq i_{end}-\mathcal{F}(n,d)+1$}
			\If{ $w_H(\vec{y}_i^{i+\mathcal{F}(n,d)-1})<d$} 
			\State remove $\vec{y}_i^{i+\mathcal{F}(n,d)-1}$ from $\vec{y}$
			\State $p(i)$: $q$-ary representation of $i$ with $\lceil\log_q n\rceil$ symbols
			\For{$j = {1,\ldots, d-1}$}		
			\If {there are at least $j$ ones in $\vec{y}_i^{i+\mathcal{F}(n,d)-1}$ }
			\State $ \begin{aligned} t(j): &q \textmd{-ary index of the } j\text{-th 1} \text{ in }\vec{y}_i^{i+\mathcal{F}(n,d)-1} \\
			&\text{with } \lceil\log_q \lceil\log_q n\rceil\rceil + C \text{ symbols} 
			\end{aligned}$
			\Else 
			\State $t(j)=1^{\lceil\log_q \lceil\log_q n\rceil\rceil + C}$
			\EndIf
			\EndFor
			\State append $p(i) t(1)\cdots t(d-1)01$ to the right of $\vec{y}$
			\State set $i_{end}=i_{end}-\mathcal{F}(n,d)$
			\State set $i=i-\mathcal{F}(n,d)+1$
			\Else
			\State set $i=i+1$
			\EndIf
			\EndWhile
		\end{algorithmic}\end{small}
	\end{algorithm}
	
The next lemma proves the correctness of Algorithm~\ref{alg:wwl-encoding}. Its proof is deferred to  Appendix~\ref{app:sec5}.
	\begin{lemma}\label{lem:encodingwwl}
		For all $n'\le n$, given any vector $\vec{x} \in \Sigma_q^{n'}$ Algorithm~\ref{alg:wwl-encoding} outputs a $(d,\mathcal{F}(n,d))$-WWL vector $\vec{y}\in \Sigma_q^{n'+d}$ such that $\vec{x}$ can be uniquely reconstructed given $\vec{y}$. The time and space complexity of the algorithm and its inverse is $\Theta(n)$.
	\end{lemma}
	
	Algorithm~\ref{alg:wwl-encoding} solves the problem of WWL encoding by replacing each subvector of small weight with the index of the subvector (denoted by $p(i)$), appended by the indices of the ones within it (denoted by $t(j)$s). The algorithm works when the window's length $k$ is $\mathcal{F}(n,d)=\log_q n + (d-1) (\log_q (\log_q n)+\Theta(1))$, with a constant number of redundancy symbols. From Lemma~\ref{lem:bounds2}, the redundancy in this case is at least $\Theta(1)$,  hence the redundancy of the algorithm is optimal up to a constant number of bits. Lastly, we can also extend this algorithm to encoding WWL vectors with smaller values of $k$, that is $k =\mathcal{F}(n,d)-f(n), \ f(n)>0$, by breaking down the length-$n$ vector to ${q^{f(n)}}$ blocks of length $n/{q^{f(n)}}$, applying Algorithm~\ref{alg:wwl-encoding} on each block, and stitching the output vectors with $d$ ones between each two subvectors. This approach yields redundancy of approximately $2dq^{f(n)}$, which is also optimal up to a constant factor, according to Lemma~\ref{lem:bounds2} and the following corollary is established
	\begin{corollary} \label{cor:wwltheta}
		Let $f(n)$ be a positive function such that $\mathcal{F}(n,d)-f(n)$ is a positive integer. The redundancy of $A_q(n,\mathcal{F}(n,d)-f(n),d)$ is $\Theta(q^{f(n)}).$
	\end{corollary}

We hereby summarize our study on the $k$-RLL and the $(d,k)$-WWL constraints. Our main results are presented in Table~\ref{tbl:const_summary}. Note that the first column of the $k$-RLL constraint lists our result for all values of $k$ such that $n-2k = \Theta(n)$ and provides the redundancy order in this case. On the other hand, the second column is targeted towards specific values of $k$ and gives an exact value of the redundancy.
\begin{small}
	\begin{table*}[t]
		\centering
		\caption{{Redundancy summary of $k$-RLL and $(d,k)$-WWL constraints}}
		\label{tbl:const_summary}
		\begin{minipage}[c]{\textwidth}
			\centering
			\begin{tabular}{|c|c|c|c|}
				\hline
				Constraint 				& $k$-RLL  & $k$-RLL  & $(d, k)$-WWL  \\
				\hline
				Redundancy  &  $\Theta(\frac{n}{q^k})$ & $\log_q e (q-1) q^{\Delta_n-z-1}$  &   $\Theta(\frac{nk^{d-1}}{q^k})$  \\
				\hline
				Comments  & $n-2k=\Theta(n)$ & $k=\lceil \log n \rceil + z, z\in \mathbb{Z}, \Delta_n= \log_q n - \lceil \log_q n \rceil $ & $k=\log n + (d-1)\log \log n - \Omega(1)$  \\
				\hline
				Stated in  & Corollary~\ref{cor:theta} & Theorem~\ref{the:main} & Corollary~\ref{cor:wwltheta}  \\
				\hline
			\end{tabular}
		\end{minipage}
	\end{table*}
\end{small}	

\section{$(d_h,d_m)$-Mutually Uncorrelated Codes}\label{sec:ECC_MUCodes}

Since substitution errors may also happen both in the writing and reading processes of DNA molecules, we impose in this section a stronger and more general version of the mutual uncorrelatedness constraint, by requiring the prefixes and suffixes to not only differ but also have a large Hamming distance. 
\begin{definition}
	A code $\cC \subseteq \Sigma^n_q$ is called a \textbf{$(d_{h}, d_{m})$-MU code} if
	\begin{enumerate}
		\item the minimum Hamming distance of the code is $d_h$,
		\item for every two not necessarily distinct words $\vec{a},\vec{b}\in \cC$, and $i\in [n-1]$: $d_H(\vec{a}_1^i, \vec{b}_{n-i+1}^n) \ge \min\{i, d_m\}$.
	\end{enumerate}
\end{definition}

We set $A_{MU}(n,q,d_h,d_m)$ to be the largest cardinality of a $(d_h, d_m)$-MU code over $\Sigma_q^n$, and $M(n,q,d)$ as the largest cardinality of a length-$n$ code over $\Sigma_q^n$ with minimum Hamming distance $d$. 
Motivated by Levenshtein's upper bound~\cite{L69}, we first provide an upper bound on the value $A_{MU}(n,q,d_h,d_m)$.
\begin{theorem} \label{thm:upper_bound_dMU}
	For all positive integers $n,q,d_h,d_m$, 
$$	A_{MU}(n,q,d_h,d_m) \le \frac{M(n,q,d)}{\lfloor{n/d_m}\rfloor},$$

	where $d=\min\{d_h,2d_m\}$. 
\end{theorem}

\begin{IEEEproof}
	Let $\cC \subseteq \Sigma_q^n$ be a $(d_h, d_m)$-MU code. We let 
	$$\widehat{\cC} = \{ (\vec{a}\vec{a})_{i+1}^{n+i} | \vec{a}\in \cC, i=\alpha\cdot d_m, \alpha\in[0,\lfloor{n/d_m}\rfloor-1] \}. $$
	That is, the code $\widehat{\cC}$ consists all cyclic shifts of words from $\cC$ by $\alpha d_m$ bits. For $\vec{a}, \vec{b}\in \cC$ let $\widehat{\vec{a}} = (\vec{a}\vec{a})_{i+1}^{n+i}, \widehat{\vec{b}} = (\vec{b}\vec{b})_{j+1}^{n+j}$ with  $i\hspace{-0.3ex}=\hspace{-0.3ex}\alpha_i d_m$ and $j\hspace{-0.3ex} =\hspace{-0.3ex}\alpha_j d_m$, $\alpha_i,\alpha_j\hspace{-0.3ex} \in \hspace{-0.3ex}[0,\lfloor{n/d_m}\rfloor\hspace{-0.3ex}-\hspace{-0.3ex}1]$. Note that $\widehat{\vec{a}}, \widehat{\vec{b}} \in \widehat{\cC}$. We prove that if $i \ne j$ or $\vec{a} \ne \vec{b},$ then $d_H(\widehat{\vec{a}}, \widehat{\vec{b}}) \ge \min\{d_h,2d_m\}$. First, if $i=j$ and $\vec{a} \ne \vec{b}$ we have that $d_H(\vec{a},\vec{b}) = d_H(\widehat{\vec{a}}, \widehat{\vec{b}})\ge d_h$. Otherwise, $i \ne j$ and we assume without loss of generality that $i > j$. Notice that $\widehat{\vec{a}}_{n-i+1}^{n-j}$ is a prefix of $\vec{a}$ and $\widehat{\vec{b}}_{n-i+1}^{n-j}$ is a suffix of $\vec{b}$. Moreover, the length of $\widehat{\vec{a}}_{n-i+1}^{n-j}$ is at least $d_m$, therefore, from the definition of $(d_h, d_m)$-MU code, $d_H(\widehat{\vec{a}}_{n-i+1}^{n-j},\widehat{\vec{b}}_{n-i+1}^{n-j}) \ge d_m$. Similarly, $\widehat{\vec{b}}_{n-j+1}^{n}\widehat{\vec{b}}_1^{n-i}$ is a prefix of $\vec{b}$ and $\widehat{\vec{a}}_{n-j+1}^{n}\widehat{\vec{a}}_1^{n-i}$ is a suffix of $\vec{a}$, thus $d_H(\widehat{\vec{a}}_{n-j+1}^{n}\widehat{\vec{a}}_1^{n-i},\widehat{\vec{b}}_{n-j+1}^{n}\widehat{\vec{b}}_1^{n-i}) \ge d_m$. Therefore, 
	$$\hspace{-0.5ex}d_H(\widehat{\vec{a}}, \widehat{\vec{b}})= d(\widehat{\vec{a}}_1^{n-i}\widehat{\vec{a}}_{n-i+1}^{n-j}\widehat{\vec{a}}_{n-j+1}^{n},\widehat{\vec{b}}_1^{i}\widehat{\vec{b}}_{n-i+1}^{n-j}\widehat{\vec{b}}_{n-j+1}^{n}) \ge 2d_m.$$ 
	We showed that $d_H(\widehat{\vec{a}}, \widehat{\vec{b}}) \ge \min\{d_h,2d_m\}$ and thus 
	$$|\widehat{\cC}| = \lfloor{n/d_m}\rfloor\cdot |\cC|\le M(n,q,d),$$ 
	where $d=\min\{d_h,2d_m\}$, and the theorem follows directly.    
\end{IEEEproof}
According to the sphere packing bound we have that $M(n,q,d)\le  {q^n}/{(Cn^{\lfloor\frac{d-1}{2}\rfloor})}$ for some constant $C$, hence the minimum redundancy of any $(d_h, d_m)$-MU code is $\left\lfloor\frac{d+1}{2}\right\rfloor \log_q n-\log_q d_m+\cO(1)$.

We are now ready to show a construction of $(d_h,d_m)$-MU codes. For the sake of simplicity, all the constructions presented in the rest of the paper are for the binary case.
	
	We say that a vector $\vec{u}\in\Sigma^{\ell}$ is a \emph{$d$-auto-cyclic vector} if for every $1\leq i\le d, \ d_H(\vec{u}, 0^i\vec u_1^{\ell-i})\ge d$. 
	In the next construction we use the following $d$-auto-cyclic vector $\vec{u}$ of length $\ell(d)=d\lceil\log d\rceil+d$, which is given by $\vec{u} = 1^{d}\vec{u}_0 \cdots \vec{u}_{\lceil\log d\rceil-1}\in \Sigma^{\ell(d)}$ such that $\vec{u}_i = ((1^{2^i}0^{2^i})^{d})_1^{d}$. For example, if $d=5$ we have $ \lceil\log d\rceil = 3$ and $\vec{u} = 11111\ 10101\ 11001\ 11110$.
	\begin{construction}\label{const:dmu}
		Let $n,k$ be two integers such that $k \geq \ell(d_m)$ and $n \geq k + \ell(d_m) + 2d_m$. Denote $n'= n-k-\ell(d_m)-2d_m$ and let $\cC_H$ be a length-$n'$ $(d_m,k)$-WWL code with minimum Hamming distance $d_h$. We define the following code.
		\begin{align*}
		\cC_2(n,k,d_h,d_m)= \{ 0^k\vec{u}1^{d_m}\vec{c}1^{d_m}\ |\  \vec{c}\in \cC_H \}.
		\end{align*}
	\end{construction}
	The correctness of Construction~\ref{const:dmu} is proved in the next theorem.
	\begin{theorem}
		The code $\cC_2(n,k,d_h,d_m)$ is a $(d_h,d_m)$-MU code.
	\end{theorem}
	\begin{IEEEproof}
		For simplicity of notation let $\cC = \cC_2(n,k,d_h,d_m)$ and $\vec{a},\vec{b}\in\cC$. The code $\cC$ has minimum distance $d_h$ since $\vec{a}_{k+\ell(d_m)+d_m+1}^{n-d_m}, \vec{b}_{k+\ell(d_m)+d_m+1}^{n-d_m}\in \cC_H$ and $\cC_H$ has minimum distance $d_h$. For the second part of the proof we use the following claim. The proof is left to the reader.
		\begin{claim} \label{clm:wwl}
			Let $\vec{x},\vec{y}$ be two $(d,k)$-WWL vectors, then the vector $\vec{x}1^d\vec{y}$ is also a $(d,k)$-WWL vector. 
		\end{claim}
	
		We show that for any $\vec{a}, \vec{b}$ which are not necessarily distinct, for all $i\in [n-1]$: $d_H(\vec{a}_1^i, \vec{b}_{n-i+1}^n) \ge \min\{i, d_m\}$. We consider the following  cases:
		\begin{enumerate}
			\item For $i\in [1,d_m]$, $\vec{a}_1^i=0^i, \vec{b}_{n-i+1}^n=1^i$, and thus $d_H(\vec{a}_1^i, \vec{b}_{n-i+1}^n) = i = \min\{i, d_m\}$. 
			\item For $i\in [d_m+1,k]$, $\vec{a}_1^i=0^i, \vec{b}_{n-i+1}^n=\vec{b}_{n-i+1}^{n-d_m} 1^{d_m}$, and hence $d_H(\vec{a}_1^i, \vec{b}_{n-i+1}^n) \ge d_m$. 
			\item For $i\in [k+1,n-d_m]$, 
			notice that $\vec{b}_{d_m+1}^n=0^{k-d_m}1^{d_m}\vec{u}_{d_m+1}^{\ell(d_m)}1^{d_m} \vec{c}1^{d_m}$, where $\vec{c}\in \cC_H$, $0^{k-d_m}$, and $\vec{u}_{d_m+1}^{\ell(d_m)}$ are all $(d_m,k)$-WWL vectors. From Claim~\ref{clm:wwl} $\vec{b}_{d_m+1}^n$ is also a $(d_m,k)$-WWL vector. Since $n-i+1 > d_m,$ $\vec{b}_{n-i+1}^n$ is a subvector $\vec{b}_{d_m+1}^n$ and as such, its first $k$ positions consist at least $d_m$ ones. Hence, $d_H(\vec{a}_1^i=0^k\vec{a}_{k+1}^i, \vec{b}_{n-i+1}^n) \ge d_m$. 
			\item For $i\in [n-d_m+1, n-1]$, let $j=n-i,$ and $\widehat{\vec{a}} = \vec{a}_1^{i}, \ \widehat{\vec{b}} = \vec{b}_{n-i+1}^n= \vec{b}_{j+1}^n $. Notice that $\widehat{\vec{a}}_{k-j}^{k-j+\ell(d_m)-1}=0^{j}\vec{u}_1^{\ell(d_m)-j}$ and $\widehat{\vec{b}}_{k-j}^{k-j+\ell(d_m)-1}=\vec{u}$. $\vec{u}$ is a $d_m$-auto-cyclic vector, hence $d_H(0^{j}\vec{u}_1^{\ell(d_m)-j}, \vec{u}) \ge d_m$ and we get $d_H(\vec{a}_1^{i}, \vec{b}_{n-i+1}^n) = d_H(\widehat{\vec{a}}, \widehat{\vec{b}})\ge d_H(\widehat{\vec{a}}_{k-j}^{k-j+\ell(d_m)}, \widehat{\vec{b}}_{k-j}^{k-j+\ell(d_m)})\ge d_m.$
			
		\end{enumerate}	
		
	\end{IEEEproof}
	
	Next, we are interested in determining the maximum value of the codes' cardinality from Construction~\ref{const:dmu}, when optimizing over all possible values of $k$. We start with the case $d_h=1$, that is, the case in which we don't require the codewords to differ by more than one symbol, but we require the MU property with distance $d_m$.
	\begin{lemma}\label{lemma:c2size}
		The redundancy  of the code $\cC_2(n,k,1,d_m)$ is minimized when $k$ is of the form $k=\log n+(d_m-1)\log\log n+\Theta(1)$. The minimal redundancy is $k+\Theta(1)=\log n+(d_m-1)\log\log n +\Theta(1)$. 
	\end{lemma}
	\begin{IEEEproof}
	The construction's redundancy includes the $k+2$ fixed bits and the redundancy bits of $\cC_H$. Therefore, values of $k$ of the form $k=\mathcal{F}(n,d_m)+\omega(1)$ result in redundancy greater than $\log n+(d_m-1)\log\log n+\Theta(1)$.
	When $k=\mathcal{F}(n,d_m)+\Theta(1)$ we can use Algorithm~\ref{alg:wwl-encoding} to construct $\cC_H$ with $\Theta(1)$ redundancy bits and get total redundancy of $\log n +(d-1)\log \log n+\Theta(1)$ bits.
	Lastly, from Corollary~\ref{cor:wwltheta}, when $k=\mathcal{F}(n,d_m)-f(n),\ f(n)\ge0$ the redundancy of $\cC_2(n,k,1,d_m)$ is $k+\Theta(2^{f(n)})=\log n +(d-1)\log \log n-f(n)+\Theta(2^{f(n)})$. This term is minimized when $f(n)=\Theta(1)$ and again, it yields total redundancy of $\log n +(d-1)\log \log n+\Theta(1)$.
\end{IEEEproof}
	
According to Lemma~\ref{lemma:c2size}, Construction~\ref{const:dmu} provides existence of $(d_h=1,d_m)$-MU codes which are $\Theta(\log \log n)$ away from the lower bound on redundancy in Theorem~\ref{thm:upper_bound_dMU}. 
Note that the results so far did not include an explicit description of an encoder and decoder for general values of $d_h, d_m$. In order to provide such an efficient construction we present in our next result a $(d_h,d_m)$-MU code with linear encoding and decoding complexities, and $\lfloor\frac{d_h+1}{2}\rfloor\log n + (d_m-1) \cdot \log \log n + \mathcal{O} (d_m\log d_m)$ redundancy bits. In this case, it is also $\Theta(\log \log n)$ away from the bound on redundancy in Theorem~\ref{thm:upper_bound_dMU}.

	For the suggested construction, we choose $k= \mathcal{F}(n,d_m)+1$ in Construction~\ref{const:dmu}, and construct the code $\cC_H$ with Algorithm~\ref{alg:wwl-encoding} to provide the WWL property. We also allow greater values of $d_h$ by incorporating a systematic code of minimum hamming distance $d_h$. This result is stated in the following corollary. 	\begin{corollary} \label{crl:dmu_eff}
		There exists a binary $(d_h,d_m)$-MU code with redundancy $\left\lfloor\frac{d_h+1}{2}\right\rfloor\log n + (d_m-1) \log\log n + \mathcal{O} (d_m \log d_m)$ and linear time and space encoding and decoding complexities.
	\end{corollary}
	
\begin{IEEEproof}[Proof sketch]
		We use Algorithm~\ref{alg:wwl-encoding} to generate a $(d_m,\mathcal{F}(n,d_m))$-WWL code of length $\tilde{n}<n$, where the choice of $\tilde{n}$ will be explained later. We then guarantee minimum distance $d_h$ by applying a systematic BCH encoder on the output from Algorithm~\ref{alg:wwl-encoding}. The approximately $\lfloor\frac{d_h-1}{2}\rfloor\log(\tilde{n})$ redundancy bits are spread within the $\tilde{n}$ bits such that the difference in the positions of each two redundancy bits is greater than $\mathcal{F}(n,d_m)+1$ (we assume that $\tilde{n}$ is large enough to allow this). That way the resulting vector is a $(d_m,\mathcal{F}(n,d_m)+1)$-WWL vector. We denote this resulting set of vectors by $\cC'$, so the code $\cC'$ is a $(d_m,\mathcal{F}(n,d_m)+1)$-WWL code with minimum distance $d_h$ and its length is a function of $\tilde{n}$ denoted by $f(\tilde{n})$. The redundancy of $\cC'$ is $\lfloor\frac{d_h-1}{2}\rfloor\log(\tilde{n}) +\mathcal{O}(1)$. We construct the code $\cC_2(n,k,d_h,d_m)$ by choosing $k=\mathcal{F}(n,d_m)+1$ and using $\cC'$ as $\cC_H$. The choice of $\tilde{n}$ is determined in a way that $f(\tilde{n})=n'=n-k-\ell(d_m)-2d_m$ is satisfied. Thus, the total redundancy of this construction is upper bounded by 
		\begin{align*}
		& \mathcal{F}(n, d_m) + \ell(d_m) + 2d_m + 1 +\left\lfloor\frac{d_h-1}{2}\right\rfloor\log n +\mathcal{O}(1) \\ 
		= & \left\lfloor\frac{d_h+1}{2}\right\rfloor\log n + (d_m-1) \log\log n + \mathcal{O} (d_m \log d_m).
		\end{align*}
	\end{IEEEproof}
	
Lastly, we note that Construction~\ref{const:dmu} improves upon Construction 3 from~\cite{Y16}, which solves the problem of $(d_h,d_m=1)$-MU codes but requires $\cO(\sqrt{n})$ redundancy bits. 

\section{MU Codes with Edit Distance}\label{MUEdit}
In this section we turn to another extension of MU codes which imposes a minimum \emph{edit distance} between prefixes and suffixes as well as on the code. This extension is motivated by several works such as~\cite{YK15} which report on deletion errors during the synthesis process of DNA strands. 

The \emph{edit distance}, denoted by $d_E(\vec{a}, \vec{b})$, of two words $\vec{a}, \vec{b}$ is the minimum number of insertions and deletions that transform $\vec{a}$ to $\vec{b}$. The \emph{minimum edit distance} of a code $\cC$ is the minimal $d$ such that for any two distinct words $\vec{a}, \vec{b} \in \cC, d_E(\vec{a}, \vec{b})\ge d$. For a word $\vec{a}$ we say that $(a_{i_1}a_{i_2}\ldots a_{i_k})$ is a subsequence of $\vec{a}$ if $1\leq i_1 < i_2<\cdots < i_k\leq n$. A common subsequence of two words $\vec{a}, \vec{b}$ is a sequence which is a subsequence of $\vec{a}$ and $\vec{b}$. We say that a sequence $\vec{c}$ is a \emph{largest common subsequence (lcs)} of $\vec{a}, \vec{b}$ if there is no common subsequence of $\vec{a}$ and $\vec{b}$ with length greater than the length of $\vec{c}$. Note that every lcs of $\vec{a}, \vec{b}$ has the same length, which will be denoted by $\ell(\vec{a},\vec{b})$. 

Next, we list several commonly known claims that will be helpful in the following proofs of this section. The claims' proofs appear in Appendix~\ref{app:sec7} for completeness. 
\begin{claim}\label{clm:ed} 
	For $\vec{a}\in \Sigma_q^n, \vec{b}\in \Sigma_q^m$: $d_E(\vec{a},\vec{b})=n+m-2\ell(\vec{a},\vec{b})$. 
\end{claim}

\begin{claim} \label{clm:ced} 
	For $\vec{a}\in \Sigma_q^n, \vec{b}\in \Sigma_q^n, \vec{c}\in \Sigma_q^m,\vec{d}\in \Sigma_q^m$: \\
	$d_E(\vec{ac},\vec{bd}) \ge \max\{d_E(\vec{a},\vec{b}), d_E(\vec{c},\vec{d})\}.$ 
\end{claim}
\begin{claim} \label{clm:red} 
	For $\vec{a}\in \Sigma_q^n, \vec{b}\in \Sigma_q^n, \vec{c}\in \Sigma_q^m, d_E(\vec{ac},\vec{b}) \ge d_E(\vec{a},\vec{b})/2. $
\end{claim}	

\begin{claim} \label{clm:zed} 
	For $\vec{a}=0^n, \vec{b}\in \Sigma_q^n, d_E(\vec{a},\vec{b}) = 2w_H(\vec{b}). $ 
\end{claim}

We are now ready to present the definition of MU codes with edit distance.
\begin{definition}\label{def:EMU}
	A code $\cC$ is called a \textbf{$(d_e, d_m)$-EMU code} if
	\begin{enumerate}
		\item the minimum edit distance of the code is $d_e$,
		\item for every two not necessarily distinct words $\vec{a},\vec{b}\in \cC$, and $i,j\in[n-1]$, if $i,j\in [d_m,n-d_m]$: $d_E(\vec{a}_1^i, \vec{b}_{n-j+1}^n) \ge d_m$,  otherwise $d_E(\vec{a}_1^i, \vec{b}_{n-j+1}^n) \ge \min\{i, j, n-i, n-j\}$.
	\end{enumerate}
\end{definition}

The second condition in Definition~\ref{def:EMU} is different from the MU constraints we introduced so far since now we require large distance of suffixes and prefixes of different lengths, while in Section~\ref{sec:ECC_MUCodes} we required large hamming distance between prefixes and suffixes of the same length. This choice of the constraint will assure that suffixes and prefixes of the addresses will not get confused even if they experienced deletions and insertions. 

We set $A_{EMU}(n,q,d_e,d_m)$ to be the largest cardinality of any $(d_e, d_m)$-EMU code over $\Sigma_q^n$, and $E(n,q,d)$ is the largest cardinality of a code over $\Sigma_q^n$ with minimum edit distance $d$. 
The following is an upper bound on the value $A_{EMU}(n,q,d_e,d_m)$. 
\begin{theorem}\label{thm:upper_bound_eMU}
	For all $n,q,d_e,d_m$, $$A_{EMU}(n,q,d_e,d_m) \le \frac{E(n,q,d)}{\lfloor{n/d_m}\rfloor},$$ where $d=\min\{d_e,d_m\}$.
\end{theorem}

\begin{IEEEproof}
		 Given a $(d_e, d_m)$-EMU code, $\cC \subseteq \Sigma_q^n$, we define $\widehat{\cC}$ and $\widehat{\vec{a}},\widehat{\vec{b}}\in \widehat{\cC}$ similarly to the proof of Theorem~\ref{thm:upper_bound_dMU}. The code $\widehat{\cC}$ is defined as the code of all cyclic shifts of $\alpha \cdot d_m$ bits of codewords of $\cC$. If $\widehat{\vec{a}},\widehat{\vec{b}}\in \widehat{\cC}$ are of the same shift, their edit distance is at least $d_e$. Otherwise, we denote $\widehat{\vec{a}} = (\vec{a}\vec{a})_{i+1}^{n+i}, \widehat{\vec{b}} = (\vec{b}\vec{b})_{j+1}^{n+j}$, where $\vec{a}, \vec{b}\in \cC$, $i,j$ are multiples of $d_m$, and we assume without loss of generality that $i > j$. The subsequence $\widehat{\vec{a}}_{n-i+1}^{n-j}$ is a prefix of $\vec{a}$ and $\widehat{\vec{b}}_{n-i+1}^{n-j}$ is a suffix of $\vec{b}$. Moreover, the length of $\widehat{\vec{a}}_{n-i+1}^{n-j}$ is at least $d_m$, therefore, from the definition of $(d_e, d_m)$-EMU codes, $d_E(\widehat{\vec{a}}_{n-i+1}^{n-j},\widehat{\vec{b}}_{n-i+1}^{n-j}) \ge d_m$. We now apply Claim ~\ref{clm:ced} twice and get that $$d_E(\widehat{\vec{a}},\widehat{\vec{b}})\ge d_E(\widehat{\vec{a}}_{n-i+1}^{n-j},\widehat{\vec{b}}_{n-i+1}^{n-j}) \ge d_m.$$
		 We showed that $d_E(\widehat{\vec{a}}, \widehat{\vec{b}}) \ge \min\{d_e,d_m\}$, thus $$|\widehat{\cC}| = \lfloor{n/d_m}\rfloor\cdot |\cC|\le E(n,q,d),$$ where $d=\min\{d_h,d_e\}$ and the theorem follows directly.    
\end{IEEEproof}
In~\cite{K13}, it was shown that 
$$E(n,q,4)\le \frac{q^n-q}{(q-1)n}$$
which aligns with the asymptotic upper bounds by~\cite{L65,T84} $E(n,q,4) \lesssim \frac{q^n}{(q-1)n}.$ 
Theorem~\ref{thm:upper_bound_eMU} therefore implies that the minimum redundancy when $\min\{d_e,d_m\}=4$ is at least $2\log_q n +\Theta(1)$.
 
The following lemma is given without a proof as it shares similar ideas with the proof of Theorem~\ref{thm:emu_const} that follows.
\begin{lemma}
	The code $\cC_2(n,k,1,d_m)$ is a $(2, d_m)$-EMU code.
\end{lemma}

Next, we slightly modify Construction~\ref{const:dmu} to allow values of $d_e$ greater than 2.	
\begin{construction}\label{const:emu}
	Let $n,k$ be two integers such that $k \geq d_m$ and $n \geq k + 2 d_m$. Let $n'= n-k-2d_m$ and $\cC_E$  be a $(d_m,k)$-WWL code of length $n'$ with minimum edit distance $d_e$. The code $\cC_3(n,k,d_e,d_m)$ is defined as follows,
	\[\cC_3(n,k,d_e,d_m)=\{0^k1^{d_m}\vec{c}1^{d_m} | \ \vec{c}\in \cC_E \}.\]
\end{construction} 

The correctness of Construction~\ref{const:emu} is proved in the next theorem. 
\begin{theorem}\label{thm:emu_const}
	The code $\cC_3(n,k,d_e,d_m)$ is a $(d_e, d_m)$-EMU code.
\end{theorem}

\begin{IEEEproof}
	We use the notation $\cC=\cC_3(n,k,d_e,d_m)$ for simplicity. Any two words $\vec{a}, \vec{b}\in \cC$ satisfy $d_E(\vec{a}_{k+d_m+1}^{n-d_m}, \vec{b}_{k+d_m+1}^{n-d_m}) \ge d_E$. Applying Claim~\ref{clm:ced} twice gives us $d_E(\vec{a}, \vec{b}) \ge d_E$.
	
	We denote $\vec{a}_1^i=\vec{x}, \vec{b}_{n-j+1}^n=\vec{y}$. Notice that a word $\vec{b} \in \cC$ has the following structure $\vec{b}_{d_m+1}^n=0^{k-d_m}1^{d_m}\vec{b}_{k+d_m+1}^{n-d_m}1^{d_m}$ such that $\vec{b}_{k+d_m+1}^{n-d_m}$ is a $(d_m,k)$-WWL vector. Therefore, according to Claim~\ref{clm:wwl}, $\vec{b}_{d_m+1}^n$ is also a $(d_m,k)$-WWL vector, and since $\vec{y}$ is a subsequence of it, $\vec{y}$ is a $(d_m,k)$-WWL vector as well.
	The following cases are considered,

	\begin{enumerate}
		\item $j\in [n-d_m], \ i=j$: in that case we show a stronger property of $\vec{x}, \vec{y}$ which claims that $d_E(\vec{x}, \vec{y}) \ge 2\min \{d_m, i, j\}$.
		If $i=j\le k$ then $\vec{x} = 0^i$ and $w_H(\vec{y})\ge \min \{d_m, i, j \}$ since $\vec{y}$ ends with $\min \{d_m, i, j \}$ 1s. From Claim~\ref{clm:zed} we get $d_E(\vec{x}, \vec{y}) \ge 2\min \{d_m, i, j \}$.
		If $i=j > k$, $\vec{x}_1^k = 0^k$ and $w_H(\vec{y}_1^k) \ge d_m$ as $\vec{y}$ is a $(d_m,k)$-WWL vector. Claim~\ref{clm:zed} yields $d_E(\vec{x}_1^k, \vec{y}_1^k)\ge 2d_m$ and $d_E(\vec{x}_1^k\vec{x}_{k+1}^i, \vec{y}_1^k\vec{y}_{k+1}^i) = d_E(\vec{x}, \vec{y}) \ge \max\{2d_m, w_H(\vec{x}_{k+1}^i,\vec{y}_{k+1}^i)\} \ge 2d_m$ according to Claim~\ref{clm:ced}.
		\item $j\in [n-d_m], \ i\ne j$ we assume that $i>j$. $\vec{x}_1^j=\vec{a}_1^j $ and $\vec{y}=\vec{b}_{n-j+1}^n$, hence $d_E(\vec{x}_1^j, \vec{y} )\ge 2\min\{d_m, i, j \}$ following the previous case. Claim~\ref{clm:red} implies $d_E(\vec{x}_1^j\vec{x}_{j+1}^i, \vec{y}) = d_E(\vec{x}, \vec{y}) \ge 2\min\{d_m, i, j \}/2= \min\{d_m, i, j \} $. The proof for $j>i$ is similar.
		\item  $j\in [n-d_m+1, n-1]$: $\vec{y}=\vec{b}_{n-j+1}^n=0^{k-(n-j)}1^{d_m}\vec{b}_{k+d_m+1}^{n-d_m}1^{d_m}$. Since $n-j<d_m$ and $\vec{b}_{k+d_m+1}^{n-d_m}$ is a $(d_m,k)$-WWL vector, we deduce that  $\vec{y}$ is a $(n-j,k)$-WWL vector. In the case of $i\ge k$, following a similar path to the proof when $j\in [n-d_m]$ leads us to $d_E(\vec{x}, \vec{y}) \ge n-j\ge \min\{i, j, n-i, n-j\}$. If $i<k$ then $j-i > d_m$ and hence $d_E(\vec{x}, \vec{y}) \ge d_m >  n-j$. 
	\end{enumerate}
\end{IEEEproof}

Even though Construction~\ref{const:emu} provides a construction of $(d_e,d_m)$-EMU codes for all $d_e$ and $d_m$ it heavily depends on the existence of codes with edit distance. The knowledge on codes with large minimum edit distance is quiet limited, in the sense that there exist codes with rate 1, however their structure is complex and there is no explicit expression for their redundancy~\cite{BGZ15}. Hence, for the rest of this section we focus on the case of edit distance four, i.e. codes correcting a single deletion or insertion. 

There exists an explicit efficient method to construct a $(d_m,\mathcal{F}(n,d_m)+1)$-WWL codes with minimum edit distance 4, which will be used as the code $\cC_E$ in Construction~\ref{const:emu}.  
For this, we use Algorithm~\ref{alg:wwl-encoding} and the well known Varshamov Tenengolts (VT) codes with edit distance four in their systematic version~\cite{AF98,VT65}.
Namely, the VT code is defined for all $n$ and $b\in [n+1]$ by 
$$VT(b) = \{ \vec{a} = (a_1,\ldots,a_n)\in \Sigma^n \ | \ \Sigma_{i=1}^ni\cdot{a_i} \equiv b (\bmod n+1)\}.$$ 

The systematic version of a VT code converts any vector of length $n'=n-\lceil\log (n+1)\rceil$ to a VT vector of length $n$ by adding $\lceil\log (n+1)\rceil$ redundancy bits in locations corresponding to powers of 2~\cite{AF98}. Any integer $i\in[0,n]$ can be represented by a sum of a subset of those indices (the subset corresponding to its binary representation). We choose to represent in those indices the integer that guarantees the fulfillment of the VT constraint by the resulting vector. 
The complexity of the encoding and decoding of this method is linear, we use it to achieve the following result.
\begin{theorem}
	There exists a construction of $(d_m,\mathcal{F}(n,d_m)+1)$-WWL code of length $n$, with minimum edit distance 4, redundancy $\log n +\mathcal{O}(d_m)$ and linear time and space encoding and decoding complexities. 
\end{theorem}

\begin{IEEEproof}[Proof sketch]
	To construct the code $\cC$ which is a $(d_m,\mathcal{F}(n,d_m)+1)$-WWL code with minimum edit distance 4, we start with a $(d_m,\mathcal{F}(n,d_m))$-WWL code $\cC_{wwl}$ of length $n'$, where $n'+2d_m+\lceil\log (n'+2d_m)\rceil=n$.	
	 An efficient algorithm for encoding and decoding of such a code was presented in Lemma~\ref{lem:encodingwwl}. We then define
	
	$$\cC_{ewwl} = \{\vec{a}_1^{i_1} 1^{d_m}\vec{a}_{i_1+1}^{i_2} 1^{d_m}\vec{a}_{i_2+1}^{n'} | \vec{a}\in\cC_{wwl}\}$$ where $i_1\le i_2$ and the choice of their values will be explained later.
	The extension to $\cC_{ewwl}$ is aimed to ensure that the final resulting code satisfies the $(d_m,\mathcal{F}(n,d_m)+1)$-WWL constraint. It is readily verified that the code 
	$\cC_{ewwl}$ is also a $(d_m,\mathcal{F}(n,d_m))$-WWL code. We now apply the systematic VT code on $\cC_{ewwl}$ to get the code $\cC$ with minimum edit distance 4. The length of $\cC_{ewwl}$ is $n' +2d_m$, hence the length of $\cC$ is $n'+2d_m+\lceil\log (n'+2d_m)\rceil=n$. We are only left with showing that $\cC$ is a $(d_m,\mathcal{F}(n,d_m)+1)$-WWL code.
	
	 Recall that the redundancy bits are located in indices which are powers of 2. As a result, for any $\vec{a}\in \cC$, $\vec{a}_{2\lceil\log n\rceil+1}^{n}$ does not include a window of length $\mathcal{F}(n,d_m)=\log n + o(\log n)$ that consists of more than one redundancy bit. Combining it with the fact that $\cC_{ewwl}$ is a $(d_m,\mathcal{F}(n,d_m))$-WWL code we get that $\vec{a}_{2\lceil\log n\rceil+1}^{n}$ is a $(d_m,\mathcal{F}(n,d_m)+1)$-WWL vector.
	
	Lastly, we can choose $i_1$ and $i_2$ when constructing the code $\cC_{ewwl}$ such that a vector
	$\vec{a}\in \cC$ satisfies $\vec{a}_{\lceil\log n\rceil-d_m}^{\lceil\log n\rceil}=1^{d_m}$ and $\vec{a}_{2\lceil\log n\rceil-d_m}^{2\lceil\log n\rceil}=1^{d_m}$, or in other words 
	 $\vec{a} = \vec{x} 1^{d_m}\vec{y} 1^{d_m}\vec{z}$ where $\vec{x}, \vec{y}$ are of length $\lceil\log n\rceil-d_m$ and $\vec{z}$ of length $n-2\lceil\log n\rceil$. In that case, we showed that $\vec{z}$ is a $(d_m,\mathcal{F}(n,d_m)+1)$-WWL vector and since the length of $\vec{x},\vec{y}$ is smaller than $\mathcal{F}(n,d_m)$ they are also $(d_m,\mathcal{F}(n,d_m)+1)$-WWL vectors. Using Claim~\ref{clm:wwl} we conclude that $\vec{a}\in \cC$ is a $(d_m,\mathcal{F}(n,d_m)+1)$-WWL vector. 
\end{IEEEproof}

We integrate the code $\cC$ from the proof above as $\cC_E$ in Construction~\ref{const:emu} and conclude the result in the following corollary. 

\begin{theorem}
	There exists a construction of $(4,d_m)$-EMU codes with redundancy $ 2 \log n + (d_m-1)\log\log n + \mathcal{O} (d_m)$ and linear time and space complexity.
\end{theorem}

To summarize, in this section we first introduced in Theorem~\ref{thm:upper_bound_eMU} a lower bound on the redundancy of $(d_e,d_m)$-EMU codes. We then presented a general structure of a $(d_e,d_m)$-EMU code in Construction~\ref{const:emu}, and used this structure to obtain an explicit construction, with efficient encoder and decoder, when $d_e=4$. For this case, the construction is $(d_m-1)\log \log n+\Theta(1)$ redundancy bits away from the lower bound in Theorem~\ref{thm:upper_bound_eMU}.

\section{Balanced Mutually Uncorrelated Codes}\label{BalancedMUCodes}
In this section we study yet another extension of MU codes. Under this setup we seek the codes to be balanced. 
A binary word of length $n$, when $n$ is even, is said to be \emph{balanced} if its Hamming weight is $n/2$. It is well known that the number of balanced words is $\binom{n}{n/2} \approx \frac{2^{n+1}}{\sqrt{2\pi n}}$. 
For the extension of $q>2$, when $q$ is even, we follow the balanced definition from~\cite{Y16} and say that a code $\cC \subseteq \Sigma_q^n$ is balanced if for any $\vec{a}\in \cC$ the number of positions $i$ such that $a_i\in [0,\frac{q}{2}-1]$ is $ n/2$. Hence, the number of $q$-ary balanced words is $\binom{n}{n/2} (q/2)^{n} \approx \frac{2q^{n}}{\sqrt{2\pi n}}$. For the rest of this section we assume that $n$ and $q$ are even. A code $\cC\subseteq \Sigma_q^n$ is said to be a \emph{balanced MU code} if $\cC$ is balanced and is also an MU code. Let $A_{BMU}(n,q)$ denote the maximum cardinality of balanced MU codes. 

\begin{theorem} \label{thm:upper_bound_bmu}
	For all $n,q$, 
	$ A_{BMU}(n,q) \le \frac{\binom{n}{n/2}\left(\frac{q}{2}\right)^{n}}{n}\approx \frac{2q^{n}}{n\sqrt{2\pi n}}.$
	In particular, the redundancy of balanced MU codes is at least $1.5\log n + \cO(1)$.
\end{theorem}

\begin{IEEEproof}
	Assume that $\cC$ is a balanced MU code and let $\widehat{\cC}$ be the following code.
	$$\widehat{\cC} = \{ (\vec{a}, \vec{a})_i^{i+n-1} \ | \ \vec{a}\in \cC, i\in[1,n] \}.$$
	That is, $\widehat{\cC}$ is the code of all cyclic shifts of words from $\cC$. The code ${\cC}$ is balanced and hence $\widehat{\cC}$ is balanced as well. Furthermore, since the code $\cC$ is an MU code, the cardinality of $\widehat{\cC}$ is $n\cdot |\cC|$, and we obtain $|\widehat{\cC}| = n\cdot |\cC| \le \binom{n}{n/2}(q/2)^{n}$. 
\end{IEEEproof}

Next we present a construction of binary balanced MU codes. 
\begin{construction}\label{const:bmu}
	Let $n,k$ be two integers such that $1\le k < n$. The code $\cC(n,k)\subseteq \Sigma^n $ is defined as follows,
	\begin{align*}
	\cC_4(n,k)=\{0^k1\vec{c}1\ |\ & w_H(\vec{c})=\frac{n}{2} -2, \\
    & \vec{c} \textmd{ has no zeros run of length $k$} \}.
	\end{align*}

\end{construction}
The correctness and redundancy result of Construction~\ref{const:bmu} are stated in the next theorem.
It follows from the proof that the redundancy of the maximal $\cC_4(n,k)$ is $1.5 \log n +\cO(1)$ and from Theorem~\ref{thm:upper_bound_bmu}, this result is optimal up to a constant number of redundancy bits.   
\begin{theorem}\label{thm:const_bmu}
	The code $\cC_4(n,k)$ is a balanced MU code, and for an integer $k= \log n + a$, $ |\cC_4(n,\log n + a)| \gtrsim C\frac{2^n}{n\sqrt{n}}$, where $C= \frac{2^a-1}{2^{2a+1}\sqrt{2\pi }}$.
\end{theorem}
\begin{IEEEproof}
	The cardinality of $\cC_4(n,k)$ is the number of binary words of length $n' = n-k-2$ which consist of $n/2-2 $ ones and do not contain a zeros run of length $k$. To lower bound the number of these words we count all words of length $n'$ and weight $n/2-2$, $\binom{n'}{ n/2-2 }$, and reduce an upper bound on the number of such words that contain a zeros run of length $k$. For any word $\vec{a}\in \Sigma^{n'}$ of weight $n/2-2$ with zeros run of length $k$ there exists an index $i\in [1, n'-k+1]$ such that $\vec{a}_i^{i+k-1}=0^k$ and the remaining $n'-k$ bits consist of exactly $n/2-2$ 1s. There are $n'-k+1$ possibilities for $i$, and for any fixed $i$, there are $\binom{n'-k}{ n/2-2 }$ possibilities for the remaining $n'-k$ symbols. Therefore, the number of vectors of weight $n/2-2$  that have a zeros run of length $k$ is upper bounded by $(n'- k + 1)\binom{n'-k}{ n/2-2 }$. Based on this observation we show that $|\cC_4(n,k)|  \gtrsim \frac{2^n}{n\sqrt{n}} \frac{2^a-1}{2^{2a+1}\sqrt{2\pi }}$. The technical details of this step appear in Appendix~\ref{app:sec8}.
\end{IEEEproof}

The extension of Construction~\ref{const:bmu} for non-binary is direct, since every symbol can store $q/2$ values after the assignment of binary values. Hence the number of redundancy symbols remains the same, i.e., $1.5\log_q n +\cO(1)$. This meets the result from~\cite{Y16} where a balanced MU code over the alphabet $\{A,C,G,T\}$ was suggested, with redundancy of $1.5\log_4 (n) + \mathcal{O}(1)$ symbols. However, our construction is also applicable for the binary case as opposed to the one in~\cite{Y16}.

Lastly, we discuss efficient implementation of Construction~\ref{const:bmu}. In~\cite{K86}, Knuth presented an efficient (linear complexity) algorithm to construct balanced words. His algorithm is based on the observation that for every binary vector $\vec{a}$ there exists an index $i\in [1,n]$ such that the vector $\bar{\vec{a}}_1^i \vec{a}_{i+1}^n$ is balanced. To convert an arbitrary vector $\vec{a}$ to a balanced vector, one can store the balanced vector $\bar{\vec{a}}_1^i \vec{a}_{i+1}^n$, and append a binary balanced representation of the index $i$. The total redundancy of this method is $\log n + \log\log n + o(\log\log n)$. We extend Knuth's method to provide an efficient construction of balanced MU codes with linear encoding and decoding complexity
and redundancy $2\log n + \log \log n + o(\log \log n)$. 
\begin{theorem}\label{thm:alg_bmu}
	There exists a construction of balanced MU codes with $2\log n + \log \log n + o(\log \log n)$ redundancy bits and linear  time and space complexity.
\end{theorem}
\begin{IEEEproof}
		We describe an algorithm for efficiently encoding words of length $n$, that agree with the structure presented in Construction~\ref{const:bmu}. We assume for simplicity that $\log n, \log \log n, \log \log \log n$ are integers and we denote by $\#_1(\vec{a}), \#_0(\vec{a})$ the number of ones and zeros in $\vec{a}$, respectively. We start with a vector $x$ of length $$n'= n - 2\log n - \log \log n - 2\log \log \log n -14$$ and apply Algorithm~\ref{alg:balanced-encoding}. 
		
	\begin{algorithm}\begin{small}
			\caption{Extended Knuth's Algorithm for balanced MU codes}\label{alg:balanced-encoding}
			\begin{algorithmic}[1]
				\vspace{.1ex}
				\Require $\vec{x}\in \Sigma^{n'}$ 
				\Ensure balanced $\vec{y}\in \Sigma^{n}$, $\vec{y}=0^{\log n+3} 1 \vec{y'}1$ where $\vec{y'}$ does not contain a run of zeros of length ${\log n+3}$.  
				\State  Execute Algorithm~\ref{alg:encoding} to remove zeros runs of length $ \log n +1$ from $\vec{x}$ \label{state1}
				\State Let  $\vec{v}\in \Sigma^{n'}, \vec{v}_i=\sum\limits_{j=1}^{i}\vec{x}_j$ \label{state2}
				\State $\vec{v} = 0^{\log n+3}1\vec{v}$ \label{state3}
				\State If $\vec{v}$ is balanced set $i=0$, otherwise find an index $i$ such that $\vec{v}_1^i\bar{\vec{v}}_{i+1}^{n'}$ is balanced.  \label{state4}
				\If {$ 0.5\log n+2< i \le \log n + 4 $}  \label{if1}
				\State $\vec{w}=\vec{v}_1^{\log n + 4} \bar{\vec{v}}_{\log n + 5}^{n'}01$ \label{state5}
				\Else \label{else1}
				\State $\vec{w}=\vec{v}10$ \label{state7}
				\EndIf
				\State If $\vec{w}$ is balanced set $i=0$, otherwise find an index $i$ such that $\vec{w}_1^i\bar{\vec{w}}_{i+1}^{n'}$ is balanced\label{state6}
				\If{$i > \log n + 4$} \label{if2}
					\State $\vec{p(i)}$: balanced binary representation of $i$ with $\log n+ \log \log n + 2\log \log \log n+2$ bits, such that it does not contain a zero run of length $\log n$ 
					\State $\vec{y}=\vec{w}_1^i1\bar{\vec{w}}_{i+1}^{n'}1\vec{p(i)}0001$ 
				\Else \ ($ i \le 0.5\log n+2 $) \label{else2}
					\State $\ell=\#_0(\vec{w})- \#_1(\vec{w})$  
					\State $\vec{y}=\vec{w}1^\ell\vec{z}11$ such that $\vec{z}11$ is balanced and $\vec{y}\in \Sigma^{n}$ \label{state10}
				\EndIf
				
			\end{algorithmic}\end{small}
		\end{algorithm}
		After Step~\ref{state1}, $\vec{x}$ has no zeros runs of length $\log n+1$. According to Step~\ref{state2}, $\vec{v}$ is the integral vector of $\vec{x}$, therefore, it does not contain a zeros or ones run of length $\log n +2$. Then, we add in Step~\ref{state3} the required prefix $0^{\log n+3}1$. We are now left with balancing the vector $\vec{v}$. For that purpose, we use Knuth's Algorithm with some adaptations that ensure the overall structure of the output vector remains as required by Construction~\ref{const:bmu}, i.e., with a prefix $0^{\log n+3}1$, followed by a sequence with no zeros run of length $\log n +3$, and ends with 1. As in Knuth's algorithm, we first find an index $i$ in $\vec{v}$ such that $\vec{v}_1^i\bar{\vec{v}}_{i+1}^{n'}$ is balanced. We consider the following cases in Step~\ref{state4}: 
		\begin{enumerate}
			\item $i>\log n+4$ :
			in Step~\ref{state7} we set $\vec{w}=\vec{v}10$ and in Step~\ref{state6}, $i>\log n+4$ is still satisfied. We then set $\vec{y}$ to $\vec{w}_1^i1\bar{\vec{w}}_{i+1}^{n'}$. Since $\vec{v}$ originally did not have a zeros or ones run of length $\log n+2$ other than its prefix, $\vec{w}$ does not contain a run of length $\log n+3$ and $\vec{y}=\vec{w}_2^i1\bar{\vec{w}}_{i+1}^{n'}$ does not contain a zeros run of length $\log n+3$. Similarly to Knuth's algorithm, we also append a balanced binary representation of $i$. Such a representation is available with $\log n + \log \log n + 2\log \log \log n$ bits. Since we additionally require it to not include a zeros run of length $\log n+3$ we insert $10$ in its $\log n$'th position to get $\vec{p(i)}$. The returned vector is appended with $0001$ for balancing purposes. 
			\item $i<\log n +4, i\ne 0$: here, we do not simply apply the same approach as in the previous case because we want to guarantee $0^{\log n +3}1$ is a prefix of $\vec{y}$. $\vec{v}_1^i\bar{\vec{v}}_{i+1}^{n'}$ is balanced, hence $$\#_0(\vec{v}_1^i\bar{\vec{v}}_{i+1}^{n'})= \#_1(\vec{v}_1^i\bar{\vec{v}}_{i+1}^{n'})$$ and since $$\#_0(\vec{v}_1^i\bar{\vec{v}}_{i+1}^{n'})=i+\#_1(\vec{v}), \#_1(\vec{v}_1^i\bar{\vec{v}}_{i+1}^{n'})=\#_0(\vec{v})-i $$ we have that  
			\begin{equation}\label{eq:2i}
				2i =\#_0(\vec{v})-\#_1(\vec{v}).
			\end{equation}
			\begin{enumerate}
				\item If $i\le0.5\log n+2$, then $\vec{w}=\vec{v}10$ and in Step~\ref{state6} $i$ remains the same. In Step~\ref{state10}, $$\ell = \#_0(\vec{w})-\#_1(\vec{w})\le \log n+4$$ and therefore we balance $\vec{y}$ by simply appending the sequence $1^\ell\vec{z}11$. \label{item:<0.5logn} 
				\item If $0.5\log n+1 < i< \log n +4$ then $\ell$ might exceed our desirable redundancy number of bits. Our alternative solution is to set $\vec{w}$ in Step~\ref{state5} to $\vec{v}_1^{\log n + 4} \bar{\vec{v}}_{\log n + 5}^{n'}01$. After Step~\ref{state5} $$\#_0(\vec{w})= \#_0(\vec{v}_1^{\log n + 4} \bar{\vec{v}}_{\log n + 5}^{n'}01)=\log n + 3 +\#_1(\vec{v}),$$ 
				$$\#_1(\vec{w})=\#_1(\vec{v}_1^{\log n + 4} \bar{\vec{v}}_{\log n + 5}^{n'}01)=\#_0(\vec{v})-\log n - 1 $$  and we have that 
				\begin{align*}
				\#_0(\vec{w})-\#_1(\vec{w})&=\log n + 3 +\#_1(\vec{v}) \\
				&-(\#_0(\vec{v})-\log n - 1 )\\
				&=2\log n +4-(\#_0(\vec{v})-\#_1(\vec{v}))\\
				&\stackrel{Eq.~(\ref{eq:2i})}=2\log n +4 -2i\\
				&<\log n+2.
				\end{align*}
				Therefore, after Step~\ref{state6}, similarly to Equation~(\ref{eq:2i}), $$2i=\#_0(\vec{w})-\#_1(\vec{w})<\log n+3$$
				and the algorithm follows the path as in case~\ref{item:<0.5logn}. 
			\end{enumerate}
		\item $i=\log n +4$ or $i=0$: in both cases $\vec{w}$ is balanced after the if clause in Step~\ref{if1}. In Step~\ref{state6} $i=0$, the if statement in Step~\ref{if2} is not satisfied and in Step~\ref{state10} $\ell=0$. We append some balanced suffix of the form $\vec{z}11$ and of proper length such that $\vec{y}=\vec{w}1^\ell\vec{z}11$ is a balanced length-$n$ vector as expected. 
		\end{enumerate}
		The vector $\vec{y}$ is uniquely decodable in the following manner. By looking at the two rightmost bits we can detect whether the if statement in Step~\ref{if2} was satisfied and reconstruct $\vec{w}$. Then, again, we look at the two rightmost bits to detect whether the if statement in Step~\ref{if1} was satisfied and reconstruct $\vec{v}$. Finally, $\vec{x}$ is derived from $~\vec{v}$ by removing the prefix $0^k1$, and computing the differences vector of $\vec{v}$.
 
\end{IEEEproof}

To conclude, we showed in Theorem~\ref{thm:upper_bound_bmu} that the problem of balanced MU codes can be solved with at least $1.5\log n + \Theta(1)$ redundancy bits. Construction~\ref{const:bmu} and Theorem~\ref{thm:const_bmu} provide existence and structure of a balanced MU code with such a redundancy. Lastly, in Theorem~\ref{thm:alg_bmu} we present an explicit construction with efficient encoding and decoding algorithms, with $2\log n + \Theta(1)$ redundancy bits.   

\section{Other related families of codes} \label{sec:other}
In this section we present two families of codes which are closely related to MU codes. Namely, \emph{comma-free codes}~\cite{G58} and \emph{prefixed synchronized codes}~\cite{G60,L04}. We discuss these codes and their connection with MU codes. 
\subsection{Comma-free codes}
Comma-free codes were studied in~\cite{G58}, motivated by the problem of word synchronization for block codes. A code $\cC\subseteq \Sigma_q^n$ is a comma-free code if for any two not necessarily distinct vectors $\vec{a}, \vec{b}\in \cC$ and $i\in [2,n]$, $(\vec{a}\vec{b})_i^{n+i-1} \not\in \cC$. 
We denote by $A_{CF}(n,q)$ the maximal cardinality of a comma-free code of length $n$ over $\Sigma_q$. It was shown in~\cite{G58} that 
$$A_{CF}(n,q)\le \frac{1}{n}\sum_{d|n}\mu(d)q^{n/d},$$
where $\mu(d)$ is the M\"{o}bius function and the sum is taken over all divisors of $n$. Later in~\cite{E65}, an optimal construction of odd length comma-free codes was introduced leading to $$A_{CF}(n,q) = \frac{1}{n}\sum_{d|n}\mu(d)q^{n/d} \approx \frac{q^n}{n}$$ for odd $n$.

Note that the comma-free property is weaker than the MU property, therefore an MU-code is a comma-free code but not vice versa; in particular, $A_{MU}(n,q)\leq A_{CF}(n,q)$. 

An extension to comma-free codes by Levenshtein~\cite{L69} states that $\cC\subseteq \Sigma_q^n$ is a $(d, \rho)$ comma-free code if for any $\vec{a}, \vec{b}, \vec{c} \in \cC$, $i\in [2,n]$, $d_H((\vec{a}\vec{b})_i^{i+n-1},\vec{c})\ge \rho$, and $\cC$ has a minimum Hamming distance $d$.

Levenshtein proved that there exists a construction of $(d, \rho)$-comma-free codes with redundancy of approximately $\lfloor\frac{d+1}{2}\rfloor \log n + c(\rho)$ bits, where $c(\rho)$ is a constant that depends on $\rho$. However, the encoding and decoding of such codes are complex. To allow efficient encoding and decoding, Levenshtein suggested in~\cite{L04} to use constructions based upon cossets of linear codes. But, in this case, it was shown in~\cite{B66} that the redundancy is at least $\sqrt{\rho n}$. Note that any $(d, \rho)$-MU code is also a $(d, \rho)$-comma-free code, so Construction~\ref{const:dmu} is a $(d, \rho)$-comma-free code and we can use the result of Corollary~\ref{crl:dmu_eff} to construct efficient $(d, \rho)$-comma-free code with significantly less redundancy, as described next.
\begin{corollary}
	There exists a construction of a $(d, \rho)$-comma-free code with efficient encoding and decoding and $\lfloor\frac{d+1}{2}\rfloor\log n + (\rho-1)\log \log n + o(\log \log n)$ redundancy bits. 
\end{corollary}
For the case of $\rho\ge d$ this construction is $\cO(\log \log n )$ away from the optimal possible redundancy according to~\cite{L69}.
The details of the construction are described in the proof of Corollary~\ref{crl:dmu_eff}.

\subsection{prefix synchronized codes}
For a set $H\subseteq\Sigma_q^m$, a prefix synchronized code $\cC_H\subseteq \Sigma_q^n$ is defined to be the set of all words $\vec{a}\in \Sigma^n_q$ such that for any $h\in H$, the word $\vec{a}h$ contains a word from $H$ only in the first and last $m$ positions~\cite{G60,L04}. Construction~\ref{const:mu} is in fact a prefix synchronized code with the set $H=\{0^k\}$. Prefix synchronized codes can be defined for any set of prefixes $H$. 
Another related problem is discussed in~\cite{L69}, namely, prefix synchronized codes with index $\rho$. A code $\cC\subseteq \Sigma^n$ is said to be prefix synchronized with a set $H\subseteq\Sigma_q^m, m\le n$ and index $\rho$ if for any $\vec{a}\in\cC, \ \vec{h}\in H, \   i\in [2,n], d_H((\vec{a}\vec{h})_i^{i+m-1},\vec{h})\ge \rho$. Levenshtein stated in~\cite{L69} that when $n$ goes to infinity, a lower bound on the redundancy of a prefix synchronized code with index $\rho$ is $\log n + (\rho - 1)\log \log n + \log \log \log n$. The next theorem provides a prefix synchronized code which is close to optimal.  
\begin{theorem}
	The code $\cC_2(n,k,1,d_m)$ is prefix synchronized with $H=\{0^k\vec{u}\}$ and $\rho = d_m$.
\end{theorem}
Hence, by Corollary~\ref{crl:dmu_eff} we provide an efficient construction to this problem with only $o(\log \log n)$ additional bits of redundancy.

\section{Conclusion}\label{sec:conc}
In this work we studied MU codes and their extension to MU codes with Hamming distance. For that purpose we looked into two interesting constraints, the $k$-RLL and the $(d,k)$-WWL constraints when $k$ is a function of the word's length, $n$. The results of this study are presented in Table~\ref{tbl:const_summary}. We then continued to additional variations of MU codes, that is MU codes with minimum Edit distance and balanced MU codes. Similar techniques can be applied to construct balanced MU codes together with minimum Hamming distance and thereby satisfy three of the constraints listed in~\cite{Y16}. The results on the variations of MU codes are summarized in Table~\ref{tbl:summary}. For each case we first give the lower bound on the redundancy, then the construction that solves this case, and finally the best redundancy we could get with linear encoding and decoding complexity.
\begin{small}
	\begin{table*}[t]
		\centering
		\caption{Redundancy summary for binary MU codes}
		\label{tbl:summary}
		\begin{minipage}[c]{\textwidth}
			\centering
			\begin{tabular}{|c|c|c|c|c|c|}
				\hline
				Property 				& MU  & $(d_h, d_m)$- MU  & $(4, d_m)$- EMU & Balanced MU   \\
				\hline
				Lower bound  &  $\log n +\log (e)$ &  $\lfloor{\frac{d+1}{2}}\rfloor \log n -\log d_m + \cO(1) $  &  $2\log n -\log d_m + \cO(1) $  &   $1.5\log n + \log \sqrt {2\pi} -1$ \\
				\hline
				Construction  			& Construction~\ref{const:mu} & Construction~\ref{const:dmu} &  Construction~\ref{const:emu}  & Construction ~\ref{const:bmu}       \\
				\hline
				Efficient  & $\lceil\log n\rceil +4$ & $\lfloor\frac{d_h+1}{2}\rfloor\log n + (d_m-1)\log \log n$ &  $2\log n + (d_m-1)\log \log n$  & $2\log n + \log \log n$          \\
				upper bound			&  & $+ \mathcal{O}(d_m \log d_m)$ & $+ \mathcal{O}(d_m)$  & $+o(\log \log n)$ \\
				\hline
				Comments  &  & $d=\min\{2d_m, d_h\}$ &  $d_m\ge 4$  &          \\
				\hline
			\end{tabular}
		\end{minipage}
	\end{table*}
\end{small}	

\section*{Acknowledgments}
The authors would like to thank Olgica Milenkovic and Ryan Gabrys for valuable discussions on DNA storage and Ron M. Roth for his contribution to the proof of Theorem~\ref{th:c1size}.


\appendices
\section{}\label{app:sec3}
	\begin{customproposition}{\ref{pro:powapprox}}
		Let $f(n),\ g(n)$ be functions such that $\lim_{n\rightarrow \infty} g(n) = 1$ and $1\le f(n) \le C$ for a constant $C$ then
		$$ f(n)^{g(n)}\approx f(n)$$
	\end{customproposition}
	
	\begin{IEEEproof}
		For any $0<\delta<1$ we choose 
		$$\delta'=\min\{-\frac{\log (1-\delta)}{\log C}, \frac{\log (1+\delta)}{\log C} \}>0 $$
		such that $1-\delta \le C^{-\delta'}$ and $C^{\delta'}\le 1+\delta$ are satisfied. There exists  $N'$ such that for every $n\ge N'$ 
		$$ 1-\delta' \le g(n)\le 1+\delta'. $$
		Therefore, for every $n\ge N'$ 
		$$ C^{-\delta'} \le f(n)^{-\delta'} \le  \frac{f(n)^{g(n)}}{f(n)} \le f(n)^{\delta'} \le C^{\delta'} $$
		and from the choice of $\delta'$ we get that for every $n\ge N'$ 
		$$ 1-\delta \le \frac{f(n)^{g(n)}}{f(n)}\le 1+\delta $$ hence
		$$\lim_{n\rightarrow \infty} \frac{f(n)^{g(n)}}{f(n)}=1$$

	\end{IEEEproof}

\section{}\label{app:sec4}
\begin{customlemma} {\ref{lem:mucapacity}}
	For $k=\lceil \log_q n\rceil + z, z\in \mathbb{Z}$ ,
	$$2^{n'E_{k,q}}\approx \frac{q^n}{n} \cdot \frac{q^{\Delta_n-z-2}}{e^{(q-1) q^{\Delta_n - z-1}}} ,$$ 
	where $\Delta_n = \log_q n -\lceil \log_q n\rceil$.
\end{customlemma}

\begin{IEEEproof}
	From Lemma \ref{lem:cap}, when $k=\lceil \log_q n\rceil + z$, and for $n$ large enough there exists a constant $C\ge 0$ such that
	$$ 1\le 2^{n'(\log q-E_{k,q})}\le 2^{n'\frac{C}{n}} \le 2^C.$$
	We use Proposition~\ref{pro:powapprox} with $f(n)= 2^{n'(\log q-E_{k,q})}$and $ g(n)= \frac{(q-1)\log e q^{-k -2}}{\log q- E_{k,q}}$ to get  
	$$2^{n'(\log q-E_{k,q})}\approx2^{n'(q-1)\log e q^{-(k-1) -2}}$$ and conclude that
	\begin{align*}
	2^{n'E_{k,q}} &= 2^{n'\log q+n'(E_{k,q}-\log q)}\\
	&\approx 2^{n'\log q-n'(q-1)\log e q^{-k-1}} \\
	&= q^{n'}e^{-n'(q-1) q^{-k-1}}.
	\end{align*}
	The term $n'(q-1) q^{-k-1}$ satisfies 		
	\begin{align*}
	& n'(q-1) q^{-k-1} =\\
	&=(n-(\lceil \log_q n\rceil + z)-2)(q-1) q^{-(\lceil \log_q n\rceil + z)-1}\\
	&=(n-(\lceil \log_q n\rceil + z)-2)(q-1) \frac{q^{\Delta_n - z-1}}{n}\\
	&=(q-1) q^{\Delta_n - z-1}+\small{o}(1).
	\end{align*}
	Finally, we conclude that
	\begin{align*}
	2^{n'E_{k,q}} &\approx q^{n'}e^{-(q-1) q^{\Delta_n - z-1}+\small{o}(1)}\\
	&\approx \frac{q^n}{n} \frac{q^{\Delta_n-z-2}}{e^{(q-1) q^{\Delta_n - z-1}}}
	\end{align*}
	where $\Delta_n = \log_q n -\lceil \log_q n\rceil$.
\end{IEEEproof}

\begin{customlemma} {\ref{lem:mumain}} 
	For $z\in \mathbb{Z}$, 
	$$|\cC_1(n,q,\lceil \log n\rceil + z)|\approx \frac{q^n}{n}\big(\frac{q-1}{q}\big)^2q^{\Delta_n-z-\log_q e (q-1) q^{\Delta_n - z-1 }}$$
	where $\Delta_n = \log_q n -\lceil \log_q n\rceil$.
\end{customlemma}

\begin{IEEEproof}
	From (\ref{eq:mucardinality}) we have
	$$1\leq \frac {a_q(n',k)} {2^{n'E_{k,q}}} \leq 1 + \frac{2q^{n'-\lceil{k/2}\rceil}}{2^{n'E_{k,q}}}.$$
	By using Lemma~\ref{lem:mucapacity} we get that for $k=\lceil \log_q n\rceil + z$
	$$\lim_{n\rightarrow \infty} \frac{2q^{n'-\lceil{k/2}\rceil}}{2^{n'E_{k,q}}} = \lim_{n\rightarrow \infty} \frac{2q^{n'-\lceil{k/2}\rceil}}{{q^n}{e^{ -q^{\Delta_n - z -1 }}}} =0,$$ 
	and conclude that 
	$$1\leq \lim_{n\rightarrow \infty} \frac {a_q(n',k)} {2^{n'E_{k,q}}} \leq \lim_{n\rightarrow \infty} 1 + \frac{2q^{n'-\lceil{k/2}\rceil}}{2^{n'E_{k,q}}} = 1.$$
	Hence, $$a_q(n',k)\approx2^{n'E_{k,q}}.$$
	Recall that $|\cC_1(n,q,k)|= (q-1)^2 a_q(n',k).$ Together with Lemma~\ref{lem:mucapacity} the result follows directly. 
\end{IEEEproof}

	\begin{customtheorem}{\ref{th:c1size}} 
		\[\cC_1(n,q) \approx \frac{q^n}{n}\cdot \big(\frac{q-1}{q}\big)^2 q^{F(\Delta_n)}\le  \frac{q^n}{n}\cdot\frac{q-1}{eq},\]
		where $\Delta_n= \log_q n-\lceil\log_q n\rceil$ and $$F(\Delta_n) = \max_{z\in \{-2,-1,0\}} \big\{\Delta_n-z - \log_q(e)(q-1)q^{\Delta_n-z-1}\big\}. $$ The inequality is tight when $n\rightarrow\infty$ over any subsequence of $n$ that satisfies $\Delta_n=-\log_q(q-1)$.
	\end{customtheorem}
\begin{IEEEproof}[Completion of the Proof of Theorem~\ref{th:c1size}]
In this part, we complete the proof of Theorem~\ref{th:c1size} by analyzing the function $F(\Delta_n)$.
Recall that 
$$f(\Delta_n,z)=\Delta_n-z-\log_q e (q-1) q^{\Delta_n - z-1 }$$ 
and the value of $z$ that achieves the only maximum of $f(\Delta_n,z)$ over the real numbers is $z_0=\log_q(q-1)-1+\Delta_n$. Since we are interested in integers, we next investigate which $z\in \mathbb{Z}$ maximizes $f(\Delta_n,z)$.
We consider the following two cases.
 \begin{enumerate}
 	\item $q=2$:
 	In this case, $z_0=\Delta_n-1,$ and therefore $\ -2 < z_0 \le -1$, so the maximal value is achieved for $z\in\{-2,-1\}$. Note that		
 	$$\frac{\partial{f}}{\partial {\Delta_n}} = 1-(q-1)q^{\Delta_n - z-1}=1-2^{\Delta_n - z-1},$$
 	and
 	$$\left.\frac{\partial{f}}{\partial {\Delta_n}}\right|_{z=-1}= 1-2^{\Delta_n}\ge 0,$$
 	$$\left.\frac{\partial{f}}{\partial {\Delta_n}}\right|_{z=-2} = 1-2^{\Delta_n+1}\le 0. $$
 	
 	Hence $f(\Delta_n, -1)$ is increasing and $f(\Delta_n, -2)$ is decreasing when $-1<\Delta_n\le0$. Moreover, they meet in $\Delta_n=\log (\ln 2) \approx -0.53$. 
 	From the analysis above we summarize that when $q=2,$ 
 	$$F(\Delta_n)=
 	\begin{cases}
 	f(\Delta_n, -2),\  for\  -1<\Delta_n\le \log (\ln 2) \\
 	f(\Delta_n, -1),\   otherwise  \\
 	\end{cases}
 	$$ 
 	
 	To understand which $\Delta_n$ achieves the maximal $F(\Delta_n)$ we can consider the only two options $\Delta_n\in \{-1,0\}$ and since we get $f(-1, -2) = f(0, -2)=1-\log e$ we can conclude that
 	$$\cC_1(n,q) \lesssim \frac{2^n}{4n}\cdot 2^{1-\log e}=\frac{2^n}{2en}.$$  
 	Lastly, when $n\rightarrow\infty$ over any subsequence of $n$ that satisfies $\lim_{n\rightarrow \infty}\Delta_n \in \{-1,0\}$, we have that
 	$$\cC_1(n,q) \approx \frac{2^n}{2en}.$$  
 	\item $q>2$:\\ 
 	For any $q>2$ and $-1<\Delta_n\le 0$ we have that $-2 < z_0 < 0$, and so the maximum over $z\in \mathbb{Z}$ is achieved by one of the options $z\in \{-2,-1,0\}$. Note that
 	\begin{align*}
 	f(\Delta_n,-1)-&f(\Delta_n,-2)\\ 
 	&=\Delta_n+1-\log_q e (q-1) q^{\Delta_n}\\
 	&- (\Delta_n+2-\log_q e (q-1) q^{\Delta_n+1})\\
 	&=-1+\log_q e (q-1)^2q^{\Delta_n}\\
 	&\ge -1+\frac{(q-1)^2}{q\ln q } > 0,
 	\end{align*}
 	where the last step holds for any $q>2$. Therefore, we are only left with maximizing over $z\in \{-1,0\}$. The functions $f(\Delta_n,-1)$ and $f(\Delta_n,0)$ meet when $\Delta_n=\delta_0=-\log_q\frac{(q-1)^2}{q\ln q}$. Since for $q>2$, $1\le \frac{(q-1)^2}{q\ln q}< q$, we have $-1<\delta_0\le 0$. In addition, $$\left.\frac{\partial{f}}{\partial {\Delta_n}}\right|_{z=0}=1-\frac{q-1}{q}\cdot q^{\Delta_n} >0,$$ 
 	hence $f(\Delta_n,0)$ is increasing for $-1<\Delta_n\le 0$. Also,   
 	$$\left.\frac{\partial{f}}{\partial {\Delta_n}}\right|_{{z=-1,\Delta_n=\delta_0}}=1-(q-1)\cdot q^{\delta_0}=1-\frac{q\ln q}{q-1}<0,$$ 
 	so $f(\Delta_n,-1)$ is decreasing at the point $\Delta_n=\delta_0$. We conclude that for $\Delta_n\le \delta_0, \ f(\Delta_n,0) \le f(\Delta_n,-1)$ and for $\Delta_n\ge \delta_0, \ f(\Delta_n,-1) \le f(\Delta_n,0)$. That is, 
 
 $$F(\Delta_n)=
 \begin{cases}
 f(\Delta_n, -1),\  for\  -1<\Delta_n\le \delta_0 \\
 f(\Delta_n, 0),\  \ \ otherwise  \\
 \end{cases}
 $$  
 	We next study that values of $\Delta_n$ that maximize the function $F(\Delta_n)$. We showed that $f(\Delta_n, 0)$ is increasing so its maximal value is achieved when $\Delta_n=0$ and it is $F(0)= f(0,0)= \frac{q-1}{q}-1$. The maximal value of $f(\Delta_n, -1)$ is achieved when $$\left.\frac{\partial{f}}{\partial {\Delta_n}}\right|_{z=-1}=0, $$ that is, when $\Delta_n=\delta_1=-\log_q(q-1)$. Since $-1\le \delta_1\le\delta_0$ we get $F(\delta_1)= f(-1,\delta_1)= 1-\log_q(e(q-1))$ and $F(\delta_1)\ge F(0)$. We are now ready to summarize and conclude that
 	$$\cC_1(n,q) \lesssim \frac{q^n}{n}\big(\frac{q-1}{q}\big)^2 \cdot q^{F(\delta_1)}=\frac{q^n}{n}\frac{q-1}{eq}$$
 	and when $n\rightarrow\infty$ over any subsequence of $n$ that satisfies $\lim_{n\rightarrow \infty}\Delta_n \in \{-1,0\}$ we have 
 	$$\cC_1(n,q) \approx \frac{q^n}{n}\frac{q-1}{eq}.$$ 
 \end{enumerate}
\end{IEEEproof}

\section{}\label{app:sec5}
\begin{customlemma} {\ref{lem:bounds2}} 
	Let $n,k,d$ be positive integers such that $d\le k\le n$. Then, there exists a constant $C>0$ such that for $n$ large enough
	$$a_q(n,k,d) \le q^{n-C\frac{(n-2k)k^{d-1}}{q^k}}.$$ 
\end{customlemma}
\begin{IEEEproof}
	We consider the set $A_q(2k,k,d)^{\lfloor\frac{n}{2k}\rfloor}$, that is, the set of vectors which are a concatenation of $\lfloor\frac{n}{2k}\rfloor$ vectors from $A_q(2k,k,d)$. We then append it with the set of all length-$\langle n\rangle_{2k}$ $q$-ary vectors. The resulting set of length-$n$ vectors is denoted by $B_q(n,k,d) =A_q(2k,k,d)^{\lfloor\frac{n}{2k}\rfloor}\Sigma_q^{\langle n\rangle_{2k}}$. 
	Note that $A_q(n,k,d) \subseteq B_q(n,k,d)$ and $$|B_q(n,k,d)|=a_q(2k,k,d)^{\lfloor\frac{n}{2k}\rfloor}q^{\langle n\rangle_{2k}}.$$ Hence,
	\begin{equation} \label{eq:ank2}
	a_q(n,k,d)\le a_q(2k,k,d)^{\lfloor\frac{n}{2k}\rfloor}q^{\langle n\rangle_{2k}}.
	\end{equation} 
	Let $b(k)$ be the number of vectors of length $2k$ with a subsequence of the form $[1,q-1]^d\vec{u}[1,q-1]^d$ where $\vec{u}$ is a vector of length $k$ and weight smaller than $d$ and $[1,q-1]^d$ corresponds to a sequence of $d$ non zero symbols. The value of $b(k)$ is given by
	\begin{align*}
	b(k) &= q^{2k-(k+2d)}(q-1)^{2d}(2k-(k+2d)+1)\sum_{i=0}^{d-1}\binom{k}{i}\\
	&=  q^{k-2d}(q-1)^{2d}(k-2d+1)\sum_{i=0}^{d-1}\binom{k}{i}.
	\end{align*}
	All those length $2k$ vectors are not included in the set $A_q(2k,k,d)$. Therefore,
	\begin{align} \label{eq:a2k2}
	a_q(2k,k,d) \le& q^{2k} - b(k)  \\ \nonumber
	\le& q^{2k} - q^{k-2d}(q-1)^{2d}(k-2d+1)\sum_{i=0}^{d-1}\binom{k}{i}.
	\end{align} 
	We denote $B=\sum_{i=0}^{d-1}\binom{k}{i}$. Note that $B=\Theta(k^{d-1})$, when $d$ is fixed and $k$ is arbitrary large. Combining inequalities (\ref{eq:ank2}) and (\ref{eq:a2k2}) we get 
	\begin{align*}
	a_q(n,k,d) & \le (q^{2k} - q^{k-2d}(q-1)^{2d}(k-2d+1)B)^{\lfloor\frac{n}{2k}\rfloor}q^{\langle n\rangle_{2k}}   \\ 	
	& = (q^{2k}(1-\frac{(k-2d+1)(q-1)^{2d}B}{q^{k+2d}}))^{\lfloor\frac{n}{2k}\rfloor}q^{\langle n\rangle_{2k}}   \\ 	
	& = q^n(1-\frac{(k-2d+1)(q-1)^{2d}B}{q^{k+2}})^{\lfloor\frac{n}{2k}\rfloor}   \\ 	
	& \stackrel{(a)}{\le}q^n(e^{-\frac{(k-2d+1)(q-1)^{2d}B}{q^{k+2}}})^{\lfloor\frac{n}{2k}\rfloor}   \\ 
	& \le q^{n-\log_q e \frac{(k-2d+1)(q-1)^{2d}B}{q^{k+2}}(\frac{n}{2k}-1)}   \\ 	
	& \stackrel{(b)}{\le} q^{n-C\frac{(n-2k)k^{d-1}}{q^k}},
	\end{align*}
	where (a) results from the inequality $1-x\le e^{-x} $ for all $x$ and (b) since there exists a constant $C$ such that for $n$ large enough the inequality holds. 	
\end{IEEEproof}	
	\begin{customlemma}{\ref{lem:encodingwwl}}
	For all $n'\le n$, given any vector $\vec{x} \in \Sigma_q^{n'}$ Algorithm~\ref{alg:wwl-encoding} outputs a $(d,\mathcal{F}(n,d))$-WWL vector $\vec{y}\in \Sigma_q^{n'+d}$ such that $\vec{x}$ can be uniquely reconstructed given $\vec{y}$. The time and space complexity of the algorithm and its inverse is $\Theta(n)$.
	\end{customlemma}

\begin{IEEEproof}
	First, we notice that according to the choice of $\mathcal{F}(n,d)$ and $C$ the length of $\vec{y}$ does not change throughout the execution of the algorithm, therefore $\vec{y}\in \Sigma^{n'+d}_q$. 
	There exists an index $1\le t \le n'$ such that the output vector $\vec{y}$ satisfies  $\vec{y} = (\vec{y}_1^t,1^{d},\vec{y}_{t+d+1}^{n'+d})$, where $\vec{y}_1^t$ is the remainder of $\vec{x}$ after removing the low weight vectors and $\vec{y}_{t+d+1}^{n'+d}$ is the list of pointers of the form $p(i)t(1)\cdots t(d-1)01$ representing the indices of the low weight subvectors and the positions of the ones inside each subvector. 
	
	To reconstruct $\vec{x}$ we first locate the index $t$ by scanning $\vec{y}$ from the right. We read the two rightmost symbols of $\vec{y}$, if they are $01$ we conclude that the following $\mathcal{F}(n,d)-2$ symbols are a pointer of the form $p(i)t(1)\cdots t(d-1)01$, we skip them and repeat that process until we encounter with two symbols $11$. We then construct the original $\vec{x}$ by inserting proper low weight vectors of length $\mathcal{F}(n,d)$ to the remainder part $\vec{y}_1^t$. The positions of the vectors we insert and the positions of the ones within them are determined according to the indices listed in the pointers part in $\vec{y}_{t+d+1}^{n'+d}$.
	
	We next show that $\vec{y}$ does not contain a vector of length $\mathcal{F}(n,d)$ of weight less than $d$.
	The remainder part $\vec{y}_1^t$ does not contain such a vector since we removed all low weight vectors within the while loop. Also, the separating part $1^d$ ensures that there is no vector of length $\mathcal{F}(n,d)$ that originates in $\vec{y}_1^t$ and ends in $\vec{y}_{t+d+1}^{n'+d}$. 
	Next we contradict the case of low weight vector in the addressing part. Recall that the structure of $\vec{y}_{t+d+1}^{n'+d}$ is a concatenation of pointers of the form $p(i)t(1)\cdots t(d-1)01$. Note that the weight of every index $p(i)$ or $t(j)$ is at least 1. Every vector of length $\mathcal{F}(n,d)$ in $\vec{y}_{t+d+1}^{n'+d}$ consists at least $d-1$ full indices (counting both $p_i$ and $t_j$s) and an additional one from the appended 01 pairs. Therefore the total weight of such vectors is at least $d$ as required.  
	
	Lastly, the algorithm's complexity is $\Theta(n)$ since the complexity of every pointer update $\Theta(\log n)$ and there are at most $n/\log n$ updating operations.
\end{IEEEproof}

\section{}\label{app:sec7}
\begin{customclaim} {\ref{clm:ed}} 
	For $\vec{a}\in \Sigma_q^n, \vec{b}\in \Sigma_q^m$: $d_E(\vec{a},\vec{b})=n+m-2\ell(\vec{a},\vec{b})$. 
\end{customclaim}
\begin{IEEEproof}
	We say a series of insertions and deletions that transforms $\vec{a}$ to $\vec{b}$ is \emph{canonic} if all of its deletions are of symbols from the original $\vec{a}$. In other words, there were no new symbols that were inserted and then deleted. To determine the edit distance of $\vec{a}$ and $\vec{b}$ we are interested in the minimal length of a series that transforms $\vec{a}$ to $\vec{b}$. We therefore imit our discussion, without loss of generality, to canonic series, because if a series is not canonic there exists a shorter equivalent canonic series to transform $\vec{a}$ to $\vec{b}$. 

	Any canonic series is associated with a common subsequence of $\vec{a}$ and $\vec{b}$ which is received by applying all the deletions in the series on $\vec{a}$. We denote this common subsequence by $\vec{x}\in\Sigma_q^\ell$. The number of deletions in the initial canonic series is $n-\ell$ and number of insertions is $m-\ell$, so the length of the series is $n+m-2\ell$. Equivalently, any common subsequence of $\vec{a}$ and $\vec{b}$, denoted by $\vec{x}\in\Sigma_q^\ell $ is associated with a canonic series of length $n+m-2\ell$ which consists deletions of symbols of $\vec{a}$ that do not belong to $\vec{x}$ followed by insertions of symbols of $\vec{b}$ that do not belong to $\vec{x}$. Thus, there exists a common subsequence of $\vec{a}$ and $\vec{b}$ of length $\ell$ if and only if there exists a canonic series from $\vec{a}$ to $\vec{b}$ of length $n+m-2\ell$. From the definition of edit distance, the claim follows.

\end{IEEEproof}
\begin{customclaim} {\ref{clm:ced}} 
	For $\vec{a}\in \Sigma_q^n, \vec{b}\in \Sigma_q^n, \vec{c}\in \Sigma_q^m,\vec{d}\in \Sigma_q^m$: \\
	$d_E(\vec{ac},\vec{bd}) \ge \max\{d_E(\vec{a},\vec{b}), d_E(\vec{c},\vec{d})\}.$ 
\end{customclaim}
\begin{IEEEproof}
	Without loss of generality we assume that $d_E(\vec{a},\vec{b})= \max\{d_E(\vec{a},\vec{b}), d_E(\vec{c},\vec{d})\}$. We set $\ell(\vec{a},\vec{b})=\ell_1, \ell(\vec{ac},\vec{bd})=\ell_2$. For any common subsequence $\vec{x}$ of $\vec{ac}, \vec{bd}$ of length $\ell\ge m$, the prefix of length $\ell-m$, $\vec{x}_1^{\ell-m}$ is a common subsequence of $\vec{a}, \vec{b}$ hence $\ell_1\ge \ell-m$. If we take $\vec{x}$ to be an lcs of $\vec{ac},\vec{bd}$ we get that $\ell_1 \ge \ell_2-m$. Combining this with Claim~\ref{clm:ed} we have \[ d_E(\vec{ac},\vec{bd}) = 2n+2m-2\ell_2 \ge 2n+2m-2\ell_1-2m= d_E(\vec{a}, \vec{b}). \]
\end{IEEEproof}	
\begin{customclaim} {\ref{clm:red}} 
	For $\vec{a}\in \Sigma_q^n, \vec{b}\in \Sigma_q^n, \vec{c}\in \Sigma_q^m, d_E(\vec{ac},\vec{b}) \ge d_E(\vec{a},\vec{b})/2. $
\end{customclaim}	

\begin{IEEEproof}
	We will show that \[\max\{d_E(\vec{a},\vec{b})-m, m\} \le d_E(\vec{ac},\vec{b}), \]  since every $m$ satisfies $ d_E(\vec{a}, \vec{b})/2 \le \max\{d_E(\vec{a},\vec{b})-m, m\}$ the claim follows directly.	 An lcs of $\vec{ac},\vec{b}$ is a subsequence of $\vec{b}$, therefore $\ell(\vec{ac},\vec{b}) \le n$ and from Claim~\ref{clm:ed}, $d_E(\vec{ac},\vec{b})=n+n+m-2\ell(\vec{ac},\vec{b})\ge2n+m-2n=m$. 
	
	We set $\ell(\vec{a},\vec{b})=\ell_1, \ell(\vec{ac},\vec{b})=\ell_2$. Note that since $\vec{a}$ and $\vec{b}$ are of the same length, $d_E(\vec{a}, \vec{b})$ is even. As in the proof of claim~\ref{clm:ced}, $\ell_1 \ge \ell_2-m$. Combining it with Claim~\ref{clm:ed} we have $d_E(\vec{ac},\vec{b}) = 2n+m-2\ell_2 \ge 2n+m-2\ell_1-2m= d_E(\vec{a}, \vec{b})-m$.
\end{IEEEproof}	
\begin{customclaim} {\ref{clm:zed}} 
	For $\vec{a}=0^n, \vec{b}\in \Sigma_q^n, d_E(\vec{a},\vec{b}) = 2w_H(\vec{b}). $ 
\end{customclaim}

\begin{IEEEproof}
	Since $\vec{a}=0^n$ the lcs of $\vec{a}, \vec{b}$ is $0^{n-w_H(\vec{b})}$ and $\ell(\vec{a}, \vec{b}) = n-w_H(\vec{b})$. From Claim~\ref{clm:ed} we get $d_E(\vec{a}, \vec{b})=2w_H(\vec{b})$. 
\end{IEEEproof}

\section{}\label{app:sec8}
\begin{customtheorem}{\ref{thm:const_bmu}}
	The code $\cC_4(n,k)$ is a balanced MU code, and for an integer $k= \log n + a$, $ |\cC_4(n,\log n + a)| \gtrsim C\frac{2^n}{n\sqrt{n}}$, where $C= \frac{2^a-1}{2^{2a+1}\sqrt{2\pi }}$.
\end{customtheorem}

\begin{IEEEproof}[Completion of the Proof of Theorem~\ref{thm:const_bmu}]
	\begin{align*}
	|\cC_4(n,k)| &\ge \binom{n'}{ n/2-2 } - (n'- k + 1)\binom{n'-k}{ n/2-2 } \\
	&=\binom{n'-k}{ n/2-2 } \left[\prod_{i=0}^{k-1}\frac{n'-i}{n'-(n/2-2)-i}-n'+k-1 \right] \\
	&=\binom{n-2k-2}{ n/2-2 } \left[\prod_{i=0}^{k-1}\frac{n-k-2-i}{n/2-k-i}-n+2k+1 \right] \\
	&\ge \binom{n-2k-2}{ n/2-2 } \left[2^k-n+2k+1 \right]\\ 
	&= \binom{n-2k-2}{ n/2-2 } n\left[\frac{2^k+2k+1}{n}-1 \right].
	\end{align*}
Note that	
	\begin{align*}
	\binom{n-2k-2}{ n/2-2 } &= \binom{n-2k-2}{ \frac{n-2k-2}{2} } \frac{\frac{n-2k-2}{2}!}{(\frac{n}{2}-2)! } \frac{\frac{n-2k-2}{2}!}{(\frac{n}{2}-2k)! } \\
	&= \binom{n-2k-2}{ \frac{n-2k-2}{2} } \frac{(\frac{n}{2}-k-1)!}{(\frac{n}{2}-2)! } \frac{(\frac{n}{2}-k-1)!}{(\frac{n}{2}-2k)! } \\
	&=\binom{n-2k-2}{ \frac{n-2k-2}{2} } \prod_{i=1}^{k-1}\frac{\frac{n}{2}-k-i}{\frac{n}{2}-1-i} \\
	&\ge \binom{n-2k-2}{ \frac{n-2k-2}{2} } \left(\frac{\frac{n}{2}-2k+1}{\frac{n}{2}}\right)^{k-1} \\
	&= \binom{n-2k-2}{ \frac{n-2k-2}{2} } \left(1- \frac{4k-2}{n}\right)^{k-1}.
	\end{align*}
For $k = \log n + a$ we get the following 
	\begin{align*}
	&= \binom{n-2k-2}{ \frac{n-2k-2}{2} } \left(1-\frac{4k-2}{n}\right)^{\frac{n}{4k-2}\frac{(4k-2)(k-1)}{n}}\\
	&\approx \binom{n-2k-2}{ \frac{n-2k-2}{2} }e^{-\frac{(4k-2)(k-1)}{n}} \\
	&\approx \binom{n-2k-2}{ \frac{n-2k-2}{2} }\\
	&\approx \frac{2^{n-2k-2+1}}{\sqrt{2\pi n-2k-2}} \\
	&\ge \frac{2^{n}}{2^{2a+1}n^2\sqrt{2\pi n}}.
	\end{align*}
	Finally, we conclude that
	\begin{align*}
	|\cC_4(n,k)| &\gtrsim \frac{2^{n}}{2^{2a+1}n^2\sqrt{2\pi n}} n\left[\frac{2^k+2k+1}{n}-1 \right]\\
	&=\frac{2^{n}}{2^{2a+1}n\sqrt{2\pi n}}\left[\frac{n2^a+2\log n +2a +1}{n}-1 \right] \\
	&\approx \frac{2^n}{n\sqrt{n}} \frac{2^a-1}{2^{2a+1}\sqrt{2\pi }}.
	\end{align*}
\end{IEEEproof}

\end{document}

%% file: mydefs.tex
\newtheorem{theorem}{Theorem}[section]
\newtheorem{lemma}[theorem]{Lemma}
\newtheorem{corollary}[theorem]{Corollary}
\newtheorem{proposition}[theorem]{Proposition}
\newtheorem{claim}[theorem]{Claim}

\newtheorem{definition}{Definition}[section]
\newtheorem{example}{Example}[section]
\newtheorem{construction}{Construction}
\newtheorem{remark}{Remark}[section]

 \newcommand{\qed}{\hfill \mbox{\raggedright \rule{.07in}{.1in}}}

\renewcommand{\bar}{\overline}

\renewcommand{\vec}[1]{\ensuremath{\textbf{#1}}}

\newcommand{\cC}{{\cal C}}

\newcommand{\cO}{{\cal O}}